\documentclass[%
 reprint,
 amsmath,amssymb,
 aps,
]{revtex4-2}

\usepackage{graphicx}
\usepackage{dcolumn}
\usepackage{bm}
\usepackage{array,multirow}
\usepackage{xcolor}

\begin{document}

\preprint{APS/123-QED}

\title{Relativistic-flying laser focus by a laser-produced parabolic plasma mirror}

\author{Tae Moon Jeong}
\email{taemoon.jeong@eli-beams.eu}
\author{Sergei V. Bulanov}%
\altaffiliation[Also at ]{Kansai Photon Science Research Institute, National Institutes for Quantum and Radiological Science and Technology, 8-1-7 Umemidai, Kizugawa-shi, Kyoto 619-0215, Japan}
\author{Petr Valenta}%
\altaffiliation[Also at ]{Faculty of Nuclear Sciences and Physical Engineering, Czech Technical University in Prague, Brehova 7, 11519 Prague, Czech Republic}
\author{Georg Korn}
\affiliation{%
 Institute of Physics of the ASCR, ELI-Beamlines, Na Slovance 2, 18221 Prague, Czech Republic
}%


\author{Timur Zh. Esirkepov}
\author{James K. Koga}
\author{Alexander S. Pirozhkov}
\author{Masaki Kando}
\affiliation{
 Kansai Photon Science Research Institute, National Institutes for Quantum and Radiological Science and Technology, 8-1-7 Umemidai, Kizugawa-shi, Kyoto 619-0215, Japan}%
\author{Stepan S. Bulanov}
\affiliation{%
Lawrence Berkeley National Laboratory, Berkeley, California 94720, USA
}%


\date{\today}

\begin{abstract}
The question of electromagnetic field intensification towards the values typical for strong field Quantum Electrodynamics is of fundamental importance. One of the most promising intensification schemes is based on the relativistic-flying mirror concept, which shows that the electromagnetic radiation reflected by the mirror will be frequency up-shifted by a factor of $4\gamma^2$ ($\gamma$ is the Lorentz factor of the mirror). In laser-plasma interactions, such a mirror travels with relativistic velocities through plasma and typically has a parabolic form, which is advantageous for light intensification. Thus, a relativistic-flying parabolic mirror reflects the counter-propagating radiation in a form of focused and flying electromagnetic wave with a high frequency. The relativistic-flying motion of the laser focus makes the electric and magnetic field distributions of the focus complicated, and the mathematical expressions describing the field distributions of the focus become of fundamental interest. We present analytical expressions describing the field distribution formed by an ideal flying mirror which has a perfect reflectance over the entire surface and wavelength range. The peak field strength of an incident laser pulse with a center wavelength of $\lambda_0$ and an effective beam radius of $w_e$ is enhanced by a factor proportional to $\gamma^3 ( w_e / \lambda_0 )$ in the relativistic limit. Electron-positron pair production is investigated in the context of invariant fields based on the enhanced electromagnetic field. The pair production rate under the relativistic-flying laser focus is modified by the Lorentz $\gamma$-factor and the beam radius-wavelength ratio ($w_e / \lambda_0$). We show that the electron-positron pairs can be created by colliding two counter-propagating relativistic-flying laser focuses in vacuum, each of which is formed when a 180 TW laser pulse is reflected by a relativistic-flying parabolic mirror with a  $\gamma = 12.2$.    
\end{abstract}

\maketitle


\section{\label{sec:level1}Introduction}

As femtosecond high-power laser technology advances \cite{Strickland1,Sung1,ELI}, the acceleration of charged particles and the generation of high-energy photons using high-power laser pulses have been extensively investigated \cite{Mourou1,Marklund1,Esarey1}. Much attention has been recently paid to the quantum electrodynamic (QED) phenomena under an ultra-strong laser field (known as the strong field QED (SF QED) \cite{Dunne1,Piazza1,Blackburn1}), including vacuum birefringence \cite{Klein1,Karbstein1,Valle1,Shen1}, photon-photon scattering \cite{Karplus1,Karplus2,Tollis1,Tollis2,Lundstrom1,Bulanov1,Koga1,Jeong1}, and electron-positron pair production via the Schwinger mechanics \cite{Schwinger1,SSBulanov1,Gonoskov1,Yu1}. An ultra-high laser intensity close to the Schwinger intensity ($10^{29}$ W/cm$^2$) is desirable for the QED study. Therefore, international efforts constructing a high-power laser facility having 100 PW or even higher power to EW power level are recently initiated \cite{SULF,XCELS,Danson1,Li1}. However, due to the very low probability for the QED event with the currently-available laser power, various sophisticated focusing schemes, such as multiple beam focusing \cite{SSBulanov1}, tight-focusing (including $\lambda^3$ focusing idea) \cite{Bahk1,Jeong2}, and $4\pi$-spherical focusing \cite{Gonoskov2,Jeong3} schemes, are proposed to maximize the laser field strength in the focal plane at a given laser power.

Since the QED event probability depends on the quantum nonlinearity parameter, $\chi_e$, defined as $\sqrt {|(F^{\mu \nu} p_\nu )^2|}/mcE_{Sch}$ \cite{Ritus1}, an approach to observe the QED phenomena with a relatively lower laser power is to use ultra-relativistic particles interacting with the laser field \cite{Burke1,Marklund2,King1,Tollis1,Baumann1,Piazza2}. Here, $F^{\mu \nu}$ is the electromagnetic field tensor, $p_{\nu}$ the momentum of the ultra-relativistic particle, and $E_{Sch}$ the Schwinger field, $m^2 c^3 / e \hbar$. Using the expression for the parameter, $\chi_e$, for describing the QED processes, one can say that when $\chi_e > 1$ in the electron rest frame the electric field exceeds the Schwinger limit. Another interesting approach, instead of using ultra-relativistic particles, is to use the laser field reflected from a relativistic flying mirror (RFM) \cite{Bulanov2,Quere1}. In this case, the laser field reflected by the RFM experiences the double Doppler effect \cite{Einstein1}, and its angular frequency and field strength are enhanced by a factor of $4\gamma^2$ in the relativistic limit of $\beta \rightarrow 1$. In \cite{Bulanov3}, it is demonstrated through particle-in-cell simulations that the focused intensity of the reflected laser pulse can exceed the conventionally focused laser intensity when a counter-propagating laser pulse is focused by relativistic-flying parabolic mirror (RFPM).

Due to the relativistic motion of the RFPM, the laser pulse focused by the RFPM travels with a relativistic speed as well, providing the relativistic-flying laser focus (RLF) and opening new regimes for SF QED studies \cite{Zhang1}. However, despite many interesting features introduced in \cite{Kando1,Bulanov4,Koga2,Esirkepov1,Mu1}, it is not yet clear how exactly the electromagnetic (EM) field of the RLF is distributed and propagates in time and space. Thus, it is of fundamental interest to obtain mathematical expressions describing the EM field distribution of the RLF and to apply the ultra-strong field for the study of the SF-QED occurring in a very small spacetime region. We should note that another concept of flying focus generated by a chromatic focusing of chirped laser pulses was recently introduced and received considerable interest \cite{Froula1}. 

In this paper, we present mathematical formulae describing three-dimensional field distributions of the RLF focused by a RFPM. When deriving the mathematical formulae for the field of RLF, two frames of reference are employed: one is the laboratory frame of reference (hereafter, laboratory frame) and the other the boosted frame of reference (hereafter, boost frame) which moves with the RLF. An incoming laser pulse in the laboratory frame is re-expressed in the boost frame through the Lorentz-transformation. And then, a focused field is calculated in the boost frame through the diffraction integral. The $4\pi$-spherical focusing scheme \cite{Jeong3} is applied to calculate the focused field since the f-number defined as the focal length divided by the beam size becomes $\ll$1 in the boost frame. A radially- or azimuthally-polarized (TM or TE mode) EM wave \cite{Jeong4} with a proper apodization function is assumed for an analytical mathematical expression under the $4\pi$-spherically focusing scheme. The focused field distribution in the boost frame is again Lorentz-transformed to reveal the flying characteristics of field distribution of the RLF in the laboratory frame.

The paper is organized as follows: The change in optical characteristics, such as wavelength, pulse duration and field strength, of a laser pulse reflected by a relativistic-flying flat mirror (RFFM) are briefly reviewed in Sec. 2. In Section 3, the mathematical formulae expressing the field distribution of the RLF reflected and focused by an RFPM is derived and discussed. The invariant fields based on Poincare invariants ($\mathcal{F}$ and $\mathcal{G}$) are calculated and used to find the pair production rate via the Schwinger mechanism \cite{Schwinger1,Dunne2}. The electron-positron pair production as an example of QED phenomena is investigated with field expressions of RLF in Section 4.

\section{Laser pulse reflected by the relativistic-flying flat mirror}
Let us first consider that a linearly-polarized (x-polarized) incident laser pulse is reflected by a RFFM travelling along the +z-axis with a speed of $v$  (or $\beta = v/c$), where $c$ is the speed of light (see Fig. 1). In a laboratory frame [$\mathcal{L}_1$, $x^\mu =(ct,-x,-y,-z)$], before the reflection, the laser pulse propagating along the -z-axis ($\vec{k} = -k\hat{z} = -\omega/c \hat{z}$) is expressed as,
\begin{eqnarray} \label{eq:1}
E(x,y,z;t) &=& \int_{-\infty}^{\infty} E_0 (x,y;\omega) e^{i\omega (t+z/c)} d\omega \nonumber \\
&=& E_0 (x,y)\int_{-\infty}^{\infty} G(\omega) e^{i\omega (t+z/c)} d\omega \nonumber \\
&=& E_p (x,y) e^{-(t+z/c)^2 / 2 \tau_G^2} e^{i\omega_0 (t+z/c)}.
\end{eqnarray}	
Here, $E_0 (x,y;\omega) = E_0 (x,y) G(\omega)$. A Gaussian spectrum, $G(\omega)=\exp [-(\omega - \omega_0 )^2 / 2\Delta \omega^2 ]$, is assumed for the laser pulse with a center frequency of $\omega_0$ and a Gaussian width ($\Delta \omega$) of the spectrum. $E_0 (x,y;\omega)$ and $E_p(x,y)$ are peak field strengths in spectral and time domains, respectively. The peak field strength, $E_p(x,y)$, in time is represented by $\sqrt{2\pi} \Delta \omega E_0 (x,y)$ and the Gaussian width in time, $\tau_G$, by $1/\Delta \omega$. So, the peak intensity, $\mathcal{I}_p$, in time can be calculated as $\Delta \omega^2 \mathcal{I} (\omega)$, with $\mathcal{I} (\omega) = c \epsilon_0 E_0 (x,y)/2$. It should be noted that the spectral bandwidth, $\Delta \omega_F$, and pulse duration, $\tau_F$, at FWHM (Full Width at Half Maximum) are given by $2 \sqrt{\ln 2} \Delta \omega$ and $2 \sqrt{\ln 2} \tau_G$, respectively. For the flat-top spatial beam profile ($E_0 (x,y)=E_0$) with a radius of $r_0$, the energy density, $\mathcal{E}_D (t)$, and the total laser pulse energy, $\mathcal{E}_T$, are given by $(1/2) \epsilon_0 E_p^2 \exp \left[ - \left( t + z/c \right)^2 / \tau_G^2 \right]$ and $(\sqrt{\pi} /2) \epsilon_0 A c \tau_G E_p^2$, respectively. Here, $A$ is the beam area given by $\pi r_0^2$. For the Gaussian [$\exp (-r^2 / 2w_0^2)$] and lowest-order Laguerre-Gaussian [$(r/w_0) \exp (-r^2 / 2w_0^2)$] beam profiles, the area, $A$, should be replaced by the effective area, $A_e=\pi w_e^2$, where $w_e$ is $w_0$ for the Gaussian beam and $w_0 \sqrt{2p + m + 1}$ for the p-th radial, m-th azimuthal Laguerre-Gaussian beam. Now, the intensity, $\mathcal{I}$, of the laser pulse defined as $\mathcal{E}_T / A_{eff}\tau_G$ becomes $(\sqrt{\pi} /2) \epsilon_0 c E_p^2$.
\begin{figure}[b]
\includegraphics[width=1\columnwidth]{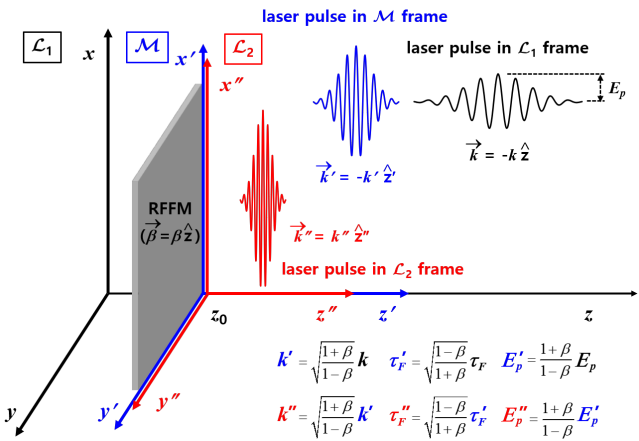}
\caption{\label{fig:epsart} Laser pulse reflected by a relativistic-flying flat mirror (RFFM). Due to the double Doppler effect, the wavevector, $k$, and the E-field strength, $E_p$, in time are enhanced by a total factor of $(1+\beta)/(1-\beta)$ and the pulse duration, $\tau_F$, is shortened by a total factor of $(1-\beta)/(1+\beta)$.}
\end{figure}

The same laser pulse can be expressed in the boost frame [$\mathcal{M}$, $x'^\mu =(ct',-x',-y',z')$] by using the Lorentz transformation. Through the Lorentz transformation, the four-vector, $x'^\mu$, and the E-field components in the boost frame are expressed as,
\begin{subequations}
\label{eq:2}
\begin{align}
        &ct' = \gamma (ct - \beta z), \: x'=x, \: y'=y, z'=\gamma(z - \beta ct), \label{eq:2-1} \\
        &\vec{E}_{\perp}' (x',y';\omega) = \gamma \left[ \vec{E}_{\perp} (x,y;\omega) + c \vec{\beta}\times \vec{B} (x,y;\omega) \right], \label{eq:2-2} \\
\intertext{and} %
        &\vec{E}_{\|}' (x',y';\omega) = \vec{E}_{\|} (x,y;\omega). \label{eq:2-3}
\end{align}
\end{subequations}
Here, the Lorentz $\gamma$-factor is defined as $1/\sqrt{1-\beta^2}$. The subscripts, $\perp$ and $\|$, refer to the polarization components perpendicular and parallel to the mirror travelling direction (+z), respectively. Since the incident laser pulse is x-polarized, $\vec{E} (x,y;\omega) = \hat{x}E_0 (x,y;\omega)$ and $\vec{B} (x,y;\omega) = -\hat{y}B_0 (x,y;\omega)$. The E-field, $E'(x',y',z';t')$, in time in the boost frame can be obtained by the Fourier transformation of $E(x',y';\omega)$ in the $\omega$ domain as,
\begin{eqnarray} \label{eq:3}
E'&&(x',y',z';t') \nonumber \\
&&= \gamma \int_{-\infty}^{\infty} \big[ E_0 (x,y;\omega) + c\beta B_0 (x,y;\omega) \big] e^{i \omega (t+z/c)} d\omega \nonumber \\
&&= \sqrt{\frac{1+\beta}{1-\beta}} E_0(x,y)\int_{-\infty}^{\infty} G(\omega) e^{i\omega \left( t+ z/c \right)} d\omega,
\end{eqnarray}	
with $E_0 = cB_0$. Then, using the Lorentz transformation given by Eq. \eqref{eq:2-1}, we obtain
\begin{eqnarray} \label{eq:4}
E'&&(x',y',z';t') \nonumber \\
&&= \sqrt{\frac{1+\beta}{1-\beta}} E_0(x',y') \int_{-\infty}^{\infty} G(\omega) e^{i \sqrt{\frac{1+\beta}{1-\beta}} \omega \left( t'+ z'/c \right)} d\omega \nonumber \\
&&= E_0(x',y')\int_{-\infty}^{\infty} G(\omega') e^{i\omega' \left( t'+ z'/c \right)} d\omega' \nonumber \\
&&= E_p ' (x,'y') e^{-(t'+z'/c)^2 / 2 {\tau_G'}^2} e^{i\omega_0 '(t'+z'/c)}.
\end{eqnarray}	
In Eq. \eqref{eq:4}, a new angular frequency, $\omega'$, defined as $\sqrt{(1+\beta)/(1-\beta)} \omega$ is introduced in the boost frame. Then, the Gaussian spectrum, $G(\omega)$, is modified as $G(\omega') = \exp [-(\omega' - \omega_0 ')^2 / 2(\Delta \omega')^2 ]$ with a new center frequency, $\omega_0'$ (=$\sqrt{(1+\beta)/(1-\beta)} \omega_0$), and spectral bandwidth, $\Delta \omega'$ (=$\Delta \omega' = \sqrt{(1+\beta)/(1-\beta)} \Delta \omega$). The peak field strength, $E_p' (x',y')$ [=$\sqrt{2\pi} \Delta \omega' E_0(x',y')$], in the boost frame is enhanced by a factor of $\sqrt{(1+\beta)/(1-\beta)}$ since $E_0(x',y')=E_0(x,y)$. The Gaussian width in time, $\tau_G' (= 1/\Delta \omega')$, is reduced by a factor of $\sqrt{(1+\beta)/(1-\beta)}$. After straightforward calculations, the following relationships for the total energy, $\mathcal{E}_T'$, and the intensity, $\mathcal{I}'$, of the laser pulse can be obtained:
\begin{equation} \label{eq:5}
\mathcal{E}_T'=\sqrt{\frac{1+\beta}{1-\beta}}\mathcal{E}_T
\quad\mathrm{and}\quad 
\mathcal{I}'=\frac{1+\beta}{1-\beta}\mathcal{I}.
\end{equation}

Next, the incident laser pulse experiences the reflection by the RFFM in the boost frame. The origin of the boost frame is located at $(0,0, z_0)$ in the laboratory frame. After the reflection in the boost frame, the propagation direction of the wavevector of the incident laser pulse is reversed ($\vec{k}' = k' \hat{z} = \omega' /c \hat{z}$). In this case, the incident laser pulse has the E-field $\vec{E}_0 ' (x',y';\omega') = \hat{x} E_0 ' (x',y';\omega')$ and the B-field $\vec{B}_0 ' (x',y';\omega') = \hat{y} B_0 ' (x',y';\omega')$. Then, from Eq. \eqref{eq:4}, the E-field of the reflected pulse is given by
\begin{eqnarray} \label{eq:6}
E_r'(x',y',z';t') &&= E_0 (x',y') \int_{-\infty}^{\infty} G(\omega') e^{i\omega' \left( t'- z'/c \right)} d\omega' \nonumber \\
&&= \int_{-\infty}^{\infty} E'(x',y';\omega') e^{i\omega' \left( t'- z'/c \right)} d\omega'.
\end{eqnarray}
The Lorentz transformations between the boost frame and another laboratory frame [$\mathcal{L}_2$, ${x''}^\mu =(ct'',-x'',-y'',-z'')$], of which the origin coincides with the boost frame, relate the four-vector and the field components as,
\begin{subequations}
\label{eq:7}
\begin{align}
&ct'' = \gamma (ct' + \beta z'), \: x''=x', \: y''=y', z''=\gamma(z' + \beta ct'), \label{eq:7-1} \\
&\vec{E}_{\perp}'' (x'',y'';\omega') = \gamma \left[ \vec{E}_{\perp}' (x',y';\omega') - c \vec{\beta}\times \vec{B}'(x',y';\omega') \right], \label{eq:7-2} \\
\intertext{and} %
        &\vec{E}_{\|}'' (x'',y'';\omega') = \vec{E}_{\|}' (x',y';\omega'). \label{eq:7-3}
\end{align}
\end{subequations}
Again, by performing the Fourier transformation into Eq. \eqref{eq:7-2} in the $\omega'$ domain, we obtain
\begin{eqnarray} \label{eq:8}
E_r'' &&(x'',y'',z'';t'')  \nonumber \\
&&= \sqrt{\frac{1+\beta}{1-\beta}} E_0 (x'',y'')  \int_{-\infty}^{\infty} G(\omega') e^{i \sqrt{\frac{1+\beta}{1-\beta}} \omega' \left( t''- z''/c \right)} d\omega' \nonumber \\
&&= E_0(x'',y'') \int_{-\infty}^{\infty} G(\omega'') e^{i \omega'' \left( t''- z''/c \right)} d\omega'' \nonumber \\
&&= E_p '' (x'',y'') e^{-(t''-z''/c)^2 / 2 {\tau_G''}^2} e^{i\omega_0 ''(t''-z''/c)}.
\end{eqnarray}	
Here, new angular frequency, $\omega''$, defined as $\sqrt{(1+\beta)/(1-\beta)}\omega' = \left[ (1+\beta)/(1-\beta) \right] \omega$ in the laboratory frame ($\mathcal{L}_2$) is introduced. So, the new center frequency, $\omega_0 ''$, and spectral bandwidth, $\Delta \omega''$, of $G(\omega'')$ in the laboratory frame ($\mathcal{L}_2$) are given by, 
\begin{equation} \label{eq:9}
\omega_0 '' = \frac{1+\beta}{1-\beta} \omega_0
\quad\mathrm{and}\quad 
\Delta \omega'' = \frac{1+\beta}{1-\beta} \Delta \omega,
\end{equation}
respectively. From Eq. \eqref{eq:8}, the peak field strength, $E_p ''(x'',y'')$, in time is again given by $\sqrt{2\pi} \Delta \omega'' E_0 (x,y)$ with $E_0 (x'',y'')=E_0 (x,y)$, yielding 
\begin{equation} \label{eq:10}
E_p '' (x'',y'')= \frac{1+\beta}{1-\beta} E_p (x,y).
\end{equation}
The Gaussian width in time, $\tau_G''$, is reduced to $1/\Delta \omega'' = \left[ (1-\beta)/(1+\beta) \right] \tau_G$. Since $t'' = t$ , $x'' = x$ , $y'' = y$, and $z'' = z - z_0$, Eq. \eqref{eq:8} can be explicitly rewritten in the original laboratory frame ($\mathcal{L}_1$) as,
\begin{eqnarray} \label{eq:11}
E_r''(t) &=& \frac{1+\beta}{1-\beta} E_p (x,y) \exp \left[ i \frac{1+\beta}{1-\beta} \omega_0 \left( t - \frac{z-z_0}{c} \right) \right] \nonumber \\
&& \times \exp \left[ -\left( \frac{1+\beta}{1-\beta} \right) ^2 \frac{1}{\tau_G^2} \left( t - \frac{z-z_0}{c} \right)^2 \right].
\end{eqnarray}	
Equation \eqref{eq:11} presents several interesting features of the laser pulse reflected from the RFFM. First, the angular frequency of the reflected pulse is enhanced by a factor of $(1+\beta)/(1-\beta)$. For instance, the center wavelength ($\lambda_0$ = 0.8 $\mu$m or 1.55 eV) of the typical PW-class Ti:S laser can be shortened to 1.24 nm (1 keV) when the Lorentz $\gamma$-factor of 12.7 [$(1+\beta)/(1-\beta) \approx 645$] is considered. Second, the pulse duration, $\tau_F''$, of the reflected pulse is shortened as $\left[ (1-\beta)/(1+\beta) \right] \tau_F$. Considering a $\gamma$-factor of 12.7 again, the pulse duration of 30 fs, which is the typical pulse duration of PW-class Ti:S laser pulse, can be reduced to 47 as. Thus, the relativistic-flying mirror with a high $\gamma$-factor can be a promising plasma optic to produce an attosecond X-ray source \cite{Pirozhkov1}.

The total energy, $\mathcal{E}_T''$, of the reflected pulse becomes $\left[ (1+\beta)/(1-\beta) \right] \mathcal{E}_T$ and its intensity, $\mathcal{I}''$, is calculated to be $\left[ (1+\beta)/(1-\beta) \right]^2 \mathcal{I}$. Thus, the total energy and the intensity of a laser pulse reflected by a RFFM are proportional to $(2\gamma)^2$ and $(2\gamma)^4$ in the relativistic limit. These basic characteristics of a laser pulse reflected by a RFFM seem very striking, since the intensity monotonically increases with the Lorentz $\gamma$-factor of the RFM and a high E-field strength above the Schwinger field is expected with a high $\gamma$-factor. However, considering that the RFM is formed by a driver laser pulse and acquires energy from the driver pulse, the total energy of the reflected pulse can be limited by the total energy, $\mathcal{E}_{DL}$, of the driver pulse. This consideration restricts the total energy of the reflected pulse as
\begin{equation} \label{eq:12}
\mathcal{E}_T'' = \frac{1+\beta}{1-\beta}\mathcal{E}_T \leq \mathcal{E}_{DL},
\end{equation}
and the highest laser intensity obtained from the RFFM is limited by $\left[ (1+\beta)(1-\beta) \right] \mathcal{E}_{DL}/A\tau_G$ when the energy of the driver laser pulse is less than $\mathcal{E}_T''$. In this case, the benefit in the intensity enhancement by the RFFM comes from the contraction in the pulse duration. Although the intensity of the laser pulse reflected by the flying flat mirror is already enhanced by a factor of $\gamma^4$ under $\mathcal{E}_{DL} \geq \mathcal{E}_T''$, the laser focus formed by an ideal RFPM provides additional enhancement factor (compared to the flat mirror case) related to the effect of frequency upshift by the double Doppler effect, so it is still of fundamental interest to derive the field expressions for the laser focus reflected by an ideal RFPM.

\section{Laser pulse reflected by the relativistic flying-parabolic mirror}
Even though the RFFM helps one understand basic properties of the reflected field, the RFPM is a more realistic plasma mirror encountered when a fs high-power laser propagates through the underdense plasma medium. A strong laser pulse (of which the normalized vector potential, $a_0$, is above unity) propagating in the plasma pushes electrons through the ponderomotive force to form a plasma cavity, and electrons return back by the recoiling force and form a high-density electron layer on the backside of the cavity. The shape of the electron layer is close to a paraboloid \cite{Esirkepov1,Sakharov1,Matlis1}, and due to the high-electron density the electron layer behaves like a parabolic mirror. Since the plasma cavity moves with a relativistic speed, the high-density electron layer forms the RFPM. A counter-propagating laser pulse is reflected and focused by the RFPM. The reflected pulse experiences the frequency upshift and the shortening of pulse duration due to the double Doppler effect as discussed in the previous section, and its focus also moves with a relativistic speed. And, when the incident laser pulse is reflected by the RFPM, due to the relativistic effect it also experiences a different curvature for the RFPM from the nominal curvature in the laboratory. Finally, all these effects related to the relativistic motion should be properly considered in calculating the field distribution of the RLF. In this work, we consider only a constantly moving mirror in the optimal regime.  
\begin{figure}[b]
\includegraphics[width=1\columnwidth]{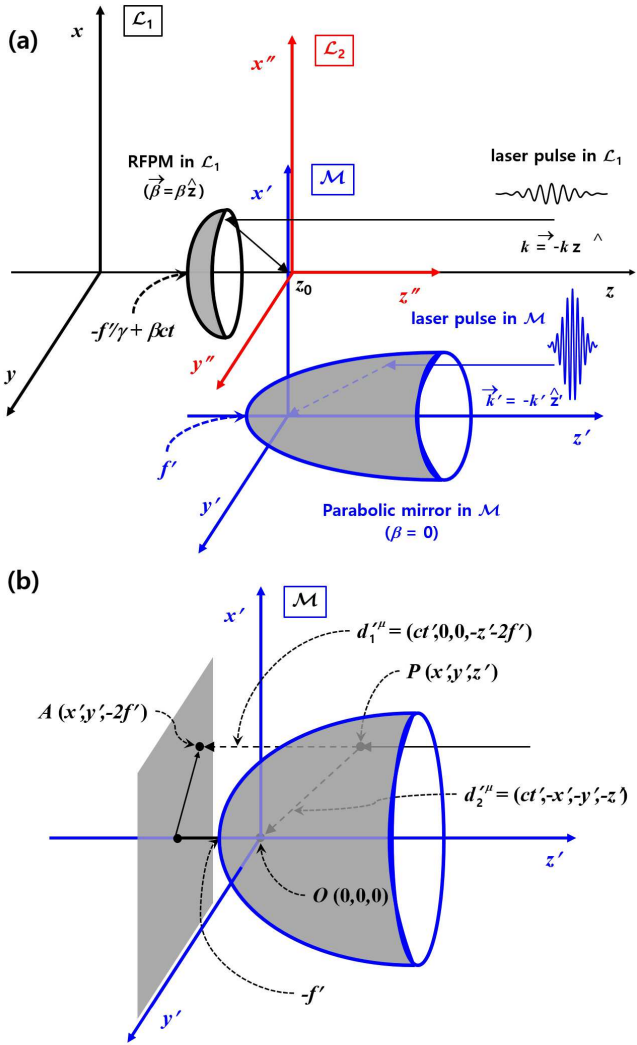}
\caption{\label{fig:epsart} Relativistic-flying parabolic mirror (RFPM) in the laboratory and the boost frames. As shown in (a), the shape of the parabolic mirror is elongated in the boost frame ($\mathcal{M}$) and the focal  length becomes short by a factor of $\gamma$ in the boost frame. The surface equation of the paraboloid in the boost frame can be obtained from the invariant property of event intervals, ${d'}_{1\mu}{d'}_1^\mu$ and ${d'}_{2\mu}{d'}_2^\mu$, shown in (b).}
\end{figure}

\subsection{\label{sec:level2}Focal length of the RFPM}
Now, in order to describe how the curvature and the focal length of the mirror change by the relativistic effect, let us first consider the equation for the surface of the RFPM. The unprimed and primed four-vectors, such as $x^\mu = (ct,-x,-y,-z)$ and ${x'}^\mu = (ct',-x',-y',-z')$, are used for describing coordinates in the laboratory frame ($\mathcal{L}_1$) and the boost frame ($\mathcal{M}$), respectively (see in Fig. 2). Assuming that the focus and the vertex of the RFPM are located at the origin and  $-f'$ on the $z'$-axis, the equation of the surface for the parabolic mirror is expressed in the boost frame as,
\begin{equation} \label{eq:13}
z'=\frac{{x'}^2 +{y'}^2}{4f'} - f',
\end{equation}
where $f'$ is known as the focal length of the RFPM. For the non-relativistic case, Eq. \eqref{eq:13} can be obtained by equating two lengths of $d_1'$ and $d_2'$ in Fig. 2(b). Here, the lengths of $d_1'$ and $d_2'$ are given by $z' + 2f'$ and $\sqrt{{x'}^2 + {y'}^2 + {z'}^2}$, respectively.

For the relativistic case, Eq. \eqref{eq:13} is generalized by the Lorentz invariant property of the interval, $x_{\mu}'{x'}^{\mu}$, between two spacetime events given by the four-vector, ${x'}^{\mu}$. When an EM wave propagates to an event $P$ from two different events ($O$ and $A$) as shown in Fig. 2(b), a four-vector between two events [$P = (0,-x',-y',-z')$ and $A = (ct',-x',-y',2f')$] is expressed by ${d'}_1^\mu = (ct',0,0,2f'+z')$ and its Lorentz-invariant interval, ${d'}_{1\mu}{d'}_1^\mu$, is given by $c^2 {t'}^2 -(2f'+z')^2$. The other four-vector between two events [$P$ and $O = (ct',0,0,0)$] is expressed by ${d'}_2^\mu = (ct',x',y',z')$ and its Lorentz-invariant interval, ${d'}_{2\mu}{d'}_2^\mu$, is $c^2 {t'}^2 - {x'}^2 - {y'}^2 - {z'}^2$. Since the RFPM moves along the  $+z$-axis, the Lorentz transformations between the laboratory and the boost frames are given as, 
\begin{subequations}
\label{eq:14}
\begin{align}
        &ct' = \gamma (ct - \beta z), \: x'=x, \: y'=y, \label{eq:14-1} \\
\intertext{and} %
        &z'=\gamma(z - \beta ct). \label{eq:14-2}
\end{align}
\end{subequations}
Now, by equating the two intervals, ${d'}_{1\mu}{d'}_1^\mu$ and ${d'}_{2\mu}{d'}_2^\mu$, we obtain the equation for the surface of the RFPM in the laboratory frame as
\begin{equation} \label{eq:15}
z=\frac{x^2 +y^2}{4\gamma f'} - \gamma f' + \frac{\gamma^2 -1}{\gamma}f' + \beta ct.
\end{equation}
In the non-relativistic limit ($\gamma \rightarrow 1$ and $\beta \rightarrow 0$), Eq. \eqref{eq:15} reduces to Eq. \eqref{eq:13}. Equation \eqref{eq:15} provides useful information on how the RFPM behaves with an incident laser pulse. First, the surface of the RFPM is described by the equation of $z=(x^2 +y^2)/4\gamma f'$ in the laboratory frame. This means that the nominal focal length ($\gamma f'$) of the RFPM in the laboratory frame is $\gamma$ times longer than that ($f'$) in the boost frame \cite{Bulanov1}. An intense ($a_0$=3) fs laser pulse propagating in a plasma medium produces a RFPM and its focal length ($\gamma f'$) observed in the laboratory frame is about 2 $\mu$m. This means that in the boost frame the focal length ($f'$) of the RFPM becomes as short as 0.1 $\mu$m with a Lorentz factor of $\gamma$=20. This is contrary to the length contraction which is well known in the special theory of relativity. The change in the focal length in the boost frame alters the focusing condition to the $4\pi$-spherical focusing scheme. For instance, the f-number (defined as the focal length divided by the beam size) changes from 0.2 to 0.01 assuming an incident beam size of $\sim$10 $\mu$m. Thus, the field distribution of the laser focus in the boost frame should be calculated under the $4\pi$-spherical focusing condition. Second, the vertex of the RFPM is located on $\gamma f' + (\gamma^2 -1) f'/\gamma + \beta ct$ at a certain time t, and its position moves with a relativistic speed of $\beta c$ in the laboratory frame. As a result, the laser focus moves with a relativistic speed of $\beta c$ in the laboratory frame as well.

For simplicity, instead of directly calculating the field distribution of the RLF in the laboratory frame, we calculate the focused field distribution first in the boost frame, and then convert it in the laboratory frame through the Lorentz transformation. 

\subsection{\label{sec:level2}Focused field in the boost frame}
Since the incident E-field in the boost frame is expressed by the Fourier transformation as in Eq. \eqref{eq:4}, a monochromatic laser field, $E'(x',y';\omega')$, in the boost frame is given by 
\begin{equation} \label{eq:16}
E'(x',y';\omega') = E_0 (x,y) G(\omega') e^{i\omega' (z'/c)}.
\end{equation}
This laser field is focused by the RFPM to form a field distribution under the $4\pi$-spherical focusing scheme in the boost frame. In this study, we assume that the incident laser pulse is radially-polarized (TM mode) or azimuthally-polarized (TE mode), since analytic solutions for those fields exist under the $4\pi$-spherically focusing condition with a specific apodization function. According to \cite{Jeong3}, in the boost frame ($\mathcal{M}$), the electric and magnetic fields of the $4\pi$-spherically focused monochromatic TM mode EM wave are expressed as,
\begin{subequations}
\label{eq:17}
\begin{align}
\vec{E}_f ' (\rho',\theta'; \omega') &= \hat{\theta'} i E_p ' (\omega ' ) a (\rho', \theta' ; \omega') e ^{i\omega' t'} \nonumber \\
&= \vec{E}_{f,\perp}' + \vec{E}_{f,\|}', \label{eq:17-1} \\
\intertext{and} %
\vec{B}_f ' (\rho',\theta'; \omega') &= -\hat{\phi'}  B_p ' (\omega ' ) b (\rho', \theta' ; \omega') e ^{i\omega' t'} \nonumber \\
&= \vec{B}_{f,\|}'. \label{eq:17-2}
\end{align}
\end{subequations}
Here, $\rho'$ (=$\sqrt{{x'}^2 + {y'}^2 + {z'}^2}$) is the magnitude of the radial displacement vector, $\vec{\rho}' = \hat{\rho}' \rho' = \hat{x}x' + \hat{y}y' + \hat{z}z'$, from the origin to an observation point $(x', y', z')$ near the origin, and $\theta'$ is the polar angle defined as $\cos^{-1} (z'/\rho')$. Again, the angular frequency, $\omega'$, in the boost frame is given by $\sqrt{(1+\beta)/(1-\beta)}\omega$ and the Gaussian spectrum in Eq. \eqref{eq:4} is assumed. The peak field strength, $E_p ' (\omega ' )$, at the focus at a certain frequency $\omega'$ is given by $(\pi/2) k' \rho_S' E_S' (\omega')$. The E-field, $E_S' (\omega')$, on a virtual sphere with a radius of $\rho_S'$ is related to the incident laser power [$P_L' (\omega') = (1/2) c \epsilon_0 {E'}^2 (x',y';\omega') A_e$] as $\sqrt{3P_L' (\omega')/4 \pi c \epsilon_0 (\rho_S')^2}$ (see Eq. (35) in \cite{Jeong3}). Thus, the peak field strength, $E_p' (\omega')$, is calculated to be  
\begin{eqnarray} \label{eq:18}
E_p' (\omega') &=& \frac{k'}{4} \sqrt{\frac{3\pi A_e}{2}} E_0 (x,y) G(\omega') \nonumber \\
 &=& \frac{k'}{4} \sqrt{\frac{3\pi A_e\mathcal{I}}{c \epsilon_0}} G(\omega') \nonumber \\
&=& C_f \mathcal{I}^{1/2} \frac{k'}{4}  G(\omega') ,
\end{eqnarray}
where $C_f$ is a constant $\sqrt{3\pi A_e /c\epsilon_0}= \sqrt{3/c\epsilon_0} \pi w_e$) related to the effective radius $w_e$, and $k'$ (= $\omega'/c$) is the magnitude of the wavevector, $\vec{k}'= \hat{\rho}'k'$, originating from the origin in the boost frame. The laser intensity, $\mathcal{I}$, in the laboratory frame is given by  $(1/2) c \epsilon_0 E_0^2 (x,y)$. 

The spatial distribution functions, $a (\rho', \theta' ; \omega')$ and $b (\rho', \theta' ; \omega')$, in Eq. \eqref{eq:17} are expressed with the $n$-th order spherical Bessel function of the first kind, $j_n (\cdot)$, and the Legendre and associated Legendre functions, $P_n (\cdot)$ and $P_n^m (\cdot)$, as
\begin{subequations}
\label{eq:19}
\begin{align}
        a \left( \rho', \theta' ; \omega' \right) =&  j_0 \left(\frac{\omega'}{c} \rho' \right) + \frac{5}{2^3} j_2 \left( \frac{\omega'}{c} \rho' \right) P_2 \left( \cos \theta' \right)  \nonumber \\
&+ \dots , \label{eq:19-1} \\
\intertext{and} %
        b \left( \rho', \theta' ; \omega'  \right) =& \frac{4}{\pi} j_1 \left( \frac{\omega'}{c} \rho'  \right) P_1^1 (\cos \theta'). \label{eq:19-2}
\end{align}
\end{subequations}
In Eq. \eqref{eq:19}, the argument, $k' \rho'$ (= $\vec{k}'\cdot \vec{\rho}'$), in spherical coordinates is replaced by $(\omega'/c)\rho'$. So, $k' \rho'$ [or $(\omega'/c)\rho'$] can be expressed as $k_x' x' +k_y' y' + k_z' z'$ in Cartesian coordinates. Since the above field distributions propagate along the z-axis in $\mathcal{L}_1$, it is convenient to express Eq. \eqref{eq:17} in Cartesian coordinates before performing the Lorentz transformation. The unit vectors, $\hat{\theta}'$ and $\hat{\phi}'$, in spherical coordinates of the boost frame are expressed as $\hat{\theta}' = \cos \theta' \cos \phi' \hat{x}' + \cos \theta' \sin \phi' \hat{y}' - \sin \theta' \hat{z}'$ and $\hat{\phi}' = \sin \phi' \hat{x}' - \cos \phi' \hat{y}'$ in Cartesian coordinates, then we re-write Eq. \eqref{eq:17} as,
\begin{subequations}
\label{eq:20}
\begin{align}
&\begin{bmatrix}
E_{f,x'}' \\ E_{f,y'}' \\ E_{f,z'}' \\
\end{bmatrix} = iE_p' (\omega') a (\rho', \theta' ; \omega' ) e^{i \omega' t'}  \begin{bmatrix}
\cos \theta' \cos \phi'  \\ \cos \theta' \sin \phi' \\ - \sin \theta' \\
\end{bmatrix}, \label{eq:20-1} \\
\intertext{and} %
&\begin{bmatrix}
B_{f,x'}' \\ B_{f,y'}' \\ B_{f,z'}' \\
\end{bmatrix} = -B_p' (\omega') b (\rho', \theta' ; \omega' ) e^{i \omega' t'}  \begin{bmatrix}
\sin \phi'  \\ -\cos \phi' \\ 0 \\
\end{bmatrix} . \label{eq:20-2}
\end{align}
\end{subequations}
The electric and magnetic fields in Eq. \eqref{eq:20} consist of in-coming ($t' + \rho'/c$) and out-going ($t' - \rho'/c$) field components. The spatial distribution function, $a (\rho', \theta' ; \omega')$, in the electric field can be approximated as $j_0 (\omega' \rho'/c)$, then $a (\rho', \theta' ; \omega') e^{i \omega' t'}$ and $b (\rho', \theta' ; \omega') e^{i \omega' t'}$ can  be separated by two parts as:
\begin{subequations}
\label{eq:21}
\begin{align}
&a (\rho', \theta' ; \omega') e^{i \omega' t'} = \left( a_+ + a_- \right) e^{i \omega' t'}, \label{eq:21-1} \\
\intertext{and} %
&b (\rho', \theta' ; \omega') e^{i \omega' t'} = \left( b_+ + b_- \right) \sin \theta' e^{i \omega' t'}, \label{eq:21-2}
\end{align}
\end{subequations}
where in-coming ($a_+$ and $b_+$) and out-going ($a_-$ and $b_-$) field components are given by,
\begin{subequations}
\label{eq:22}
\begin{align}
&a_+ = \frac{e^{i \omega' \rho'/c}}{2i \omega' \rho'/c}, \quad
a_- = -\frac{e^{-i \omega' \rho'/c}}{2i \omega' \rho'/c}, \label{eq:22-1} \\
&b_+ = \frac{4}{\pi} \left[\frac{e^{i \omega' \rho'/c}}{2i (\omega' \rho'/c)^2} - \frac{e^{i \omega' \rho'/c}}{2 \omega' \rho'/c} \right], \label{eq:22-2} \\
\intertext{and} %
&b_- = \frac{4}{\pi} \left[-\frac{e^{-i \omega' \rho'/c}}{2i (\omega' \rho'/c)^2} - \frac{e^{-i \omega' \rho'/c}}{2 \omega' \rho'/c} \right]. \label{eq:22-3}
\end{align}
\end{subequations}

In this subsection, the electromagnetic field focused by the RFPM is expressed in the boost frame. In following subsections, the Lorentz transformation of the field into the laboratory frame ($\mathcal{L}_2$) and the spatio-temporal field distribution in the laboratory frame will be explained.

\subsection{\label{sec:level2} Lorentz transformation for in-coming and out-going fields}
For phase factors of in-coming and out-going fields, the Lorentz transformations from the boost frame to the laboratory frame yield
\begin{subequations}
\label{eq:23}
\begin{align}
\omega' \left( t' +\frac{\rho'}{c} \right)  &= \omega' t' + k_x' x' +k_y' y' + k_z' z' \nonumber \\
     &= \omega' t \gamma (1 - \beta \cos \theta') + k' x \sin \theta' \cos \phi' \nonumber \\
 &+ k' y \sin \theta' \sin \phi' + k' z \gamma ( \cos \theta' - \beta), \label{eq:23-1} \\
\intertext{and} %
\omega' \left( t' -\frac{\rho'}{c} \right)  &= \omega' t' - k_x' x' -k_y' y' - k_z' z' \nonumber \\
     &= \omega' t \gamma (1 + \beta \cos \theta') - k' x \sin \theta' \cos \phi' \nonumber \\
 & - k' y \sin \theta' \sin \phi' - k' z \gamma ( \cos \theta' + \beta). \label{eq:23-2} 
\end{align}
\end{subequations}
with the expression of $(\omega'/c)\rho' = k_x' x' +k_y' y' + k_z' z'$. By introducing new variables, $\omega_+''$ and $\omega_-''$, for the angular frequencies of in-coming and out-going fields in the laboratory frame, we define 
\begin{equation} \label{eq:24}
\omega_+'' = \omega' \gamma (1-\beta \cos \theta_+') \; \mathrm{and}\; \omega_-'' = \omega' \gamma (1+\beta \cos \theta_-'),
\end{equation}
and obtain the Lorentz invariant properties for the phase as $\omega' ( t' + \rho'/c) = \omega_+'' ( t + \rho/c)$ and $\omega' ( t' - \rho'/c) = \omega_-'' ( t - \rho/c)$. Here, + and – symbols in the subscript are used to represent in-coming and out-going fields. From the Lorentz invariant properties of the phase, the following relationships between the polar angles for the in-coming and out-going fields are obtained:
\begin{subequations}
\label{eq:25}
\begin{align}
\sin \theta_+' = \frac{\sin \theta}{\gamma (1+\beta \cos \theta)} \; \mathrm{and}\; \cos \theta_+' = \frac{\cos \theta + \beta}{1+\beta \cos \theta}, \label{eq:25-1} \\
\sin \theta_-' = \frac{\sin \theta}{\gamma (1-\beta \cos \theta)} \; \mathrm{and}\; \cos \theta_-' = \frac{\cos \theta - \beta}{1-\beta \cos \theta}. \label{eq:25-2} 
\end{align}
\end{subequations}
And, with the help of Eq. \eqref{eq:25}, Eq. \eqref{eq:23} can be rewritten as,
\begin{subequations}
\label{eq:26}
\begin{align}
\omega' \left( t' +\frac{\rho'}{c} \right)  &= \frac{\omega' t}{\gamma (1+\beta \cos \theta )} +  \frac{k'x \sin \theta \cos \phi}{\gamma (1+\beta \cos \theta )} \nonumber \\
 &+  \frac{k'y \sin \theta \sin \phi}{\gamma (1+\beta \cos \theta )} +  \frac{k'z \cos \theta}{\gamma (1+\beta \cos \theta )}, \label{eq:26-1} \\
\intertext{and} %
\omega' \left( t' -\frac{\rho'}{c} \right)  &= \frac{\omega' t}{\gamma (1-\beta \cos \theta )} - \frac{k'x \sin \theta \cos \phi}{\gamma (1-\beta \cos \theta )} \nonumber \\
 &-  \frac{k'y \sin \theta \sin \phi}{\gamma (1-\beta \cos \theta )} -  \frac{k'z \cos \theta}{\gamma (1-\beta \cos \theta )}. \label{eq:26-2} 
\end{align}
\end{subequations}
So, it is clear that $\omega_\pm''$ should be expressed as 
\begin{equation} \label{eq:27}
\omega_\pm'' = \frac{\omega'}{\gamma (1 \pm \beta \cos \theta)}.
\end{equation}
in the laboratory frame. When $\theta=0$ (+z-direction), the angular frequencies for in-coming and out-going fields become $\omega_\pm'' = [(1+\beta)/(1 \pm \beta)] \omega$. Now, it is convenient to introduce new variables, $\Omega_{1,2}$ and $\Gamma_{1,2}$, defined as
\begin{subequations}
\label{eq:28}
\begin{align}
&\Omega_1 = \frac{\omega'}{\gamma (1 - \beta^2 \cos^2 \theta)}, \quad
\Omega_2 = \frac{\omega' \beta \cos \theta}{\gamma (1 - \beta^2 \cos^2 \theta)}, \label{eq:28-1} \\
&\Gamma_1 = \frac{\omega'/c}{\gamma (1 - \beta^2 \cos^2 \theta)}, \; \mathrm{and}\;
\Gamma_2 = \frac{(\omega'/c) \beta \cos \theta}{\gamma (1 - \beta^2 \cos^2 \theta)}. \label{eq:28-2}
\end{align}
\end{subequations}
Then, the phase factors for in-coming and out-going fields can be rewritten as,
\begin{subequations}
\label{eq:29}
\begin{align}
\omega' \left( t' +\frac{\rho'}{c} \right)  &= (\Omega_1 - \Omega_2)t + (\Gamma_1 - \Gamma_2) \rho, \label{eq:29-1} \\
\intertext{and} %
\omega' \left( t' -\frac{\rho'}{c} \right)  &= (\Omega_1 + \Omega_2)t - (\Gamma_1 + \Gamma_2) \rho. \label{eq:29-2} 
\end{align}
\end{subequations}
By adding or subtracting Eqs. \eqref{eq:29-1} and \eqref{eq:29-2}, we obtain
\begin{subequations}
\label{eq:30}
\begin{align}
&t' = T(t,\rho) = \frac{t-(\rho/c) \beta \cos \theta}{\gamma (1- \beta^2 \cos^2 \theta )}, \label{eq:30-1} \\
&\rho' = R(\rho,t) = \frac{\rho-ct \beta \cos \theta}{\gamma (1- \beta^2 \cos^2 \theta )},\label{eq:30-2} \\
\intertext{and} %
&t' \pm \frac{\rho'}{c}=T(t,\rho) \pm \frac{R(\rho,t)}{c} = \frac{t \pm (\rho/c)}{\gamma (1 \pm \beta \cos \theta)}. \label{eq:30-3} 
\end{align}
\end{subequations}
Equation \eqref{eq:30} shows how lightcone variables in the boost frame are Lorentz-tranformed into the laboratory frame. Hereafter, the variables, $T$ and $R$, will be used as short expressions for representing $T(t,\rho)$ and $R(\rho,t)$. 

The Lorentz transformations for the full electric and magnetic fields from the boost frame to the laboratory frame ($\mathcal{L}_2$) are given by
\begin{subequations}
\label{eq:31}
\begin{align}
&\vec{E}_{\|}'' = \vec{E}_{\|}', \quad \vec{B}_{\|}'' = \vec{B}_{\|}', \label{eq:31-1} \\
&\vec{E}_{\perp}'' = \gamma ( \vec{E}_{\perp}' - c \vec{\beta} \times \vec{B}_{\perp}'), \label{eq:31-2} \\
\intertext{and} %
&\vec{B}_{\perp}'' = \gamma ( \vec{B}_{\perp}' + \vec{\beta} \times \vec{E}_{\perp}'/c).  \label{eq:31-3} 
\end{align}
\end{subequations}
Since the parallel polarization components for the field remain unchanged through the Lorentz transformation, we have 
\begin{subequations}
\label{eq:32}
\begin{align}
E_{f,z'',\pm}'' &= E_{f,z',\pm}' \nonumber \\
&= -iE_p' (\omega') a_{\pm} \sin \theta_\pm', \label{eq:32-1} \\
\intertext{and} %
B_{f,z'',\pm}'' &= B_{f,z',\pm}'=0.  \label{eq:32-2} 
\end{align}
\end{subequations}
The perpendicular components for the focused fields in the laboratory frame are expressed as, 
\begin{subequations}
\label{eq:33}
\begin{align}
&\begin{bmatrix}
E_{f,x'',\pm}'' (\omega') \\ E_{f,y'',\pm}'' (\omega') \\
\end{bmatrix} = \gamma \begin{bmatrix} 
E_{f,x',\pm}' + c \beta B_{f,y',\pm}' \\ E_{f,y',\pm}' - c \beta B_{f,x',\pm}' \\
\end{bmatrix}  \nonumber \\
&= \gamma E_p' (\omega') \left( ia_{\pm} \cos \theta_\pm + \beta b_{\pm} \sin \theta_\pm \right)  
\begin{bmatrix}
\cos \phi' \\ \sin \phi' \\
\end{bmatrix}, \label{eq:33-1} \\
\intertext{and} %
&\begin{bmatrix}
B_{f,x'',\pm}'' (\omega') \\ B_{f,y'',\pm}'' (\omega') \\
\end{bmatrix} = \gamma \begin{bmatrix} 
B_{f,x'',\pm}' - (\beta/c) E_{f,y',\pm}' \\ B_{f,y',\pm}' + (\beta/c) E_{f,x',\pm}' \\
\end{bmatrix} \nonumber \\
&= \frac{\gamma}{c} E_p' (\omega') \left( b_{\pm} \sin \theta_\pm + i \beta a_{\pm} \cos \theta_\pm \right)  
\begin{bmatrix}
-\sin \phi' \\ \cos \phi' \\
\end{bmatrix}. \label{eq:33-2}
\end{align}
\end{subequations}
The final field expression can be obtained by summing in-coming and out-going fields as
\begin{subequations}
\label{eq:34}
\begin{align}
\begin{bmatrix}
E_{f,x''}'' (\omega') \\ E_{f,y''}'' (\omega') \\
\end{bmatrix} &= \gamma \begin{bmatrix} 
E_{f,x'}' + c \beta B_{f,y'}' \\ E_{f,y'}' - c \beta B_{f,x'}' \\
\end{bmatrix}  \nonumber \\
&= \gamma E_p' (\omega') (ia'  + \beta b' )  
\begin{bmatrix}
\cos \phi' \\ \sin \phi' \\
\end{bmatrix}, \label{eq:34-1} \\
\begin{bmatrix}
B_{f,x''}'' (\omega') \\ B_{f,y''}'' (\omega') \\
\end{bmatrix} &= \gamma \begin{bmatrix} 
B_{f,x''}' - (\beta/c) E_{f,y'}' \\ B_{f,y'}' + (\beta/c) E_{f,x'}' \\
\end{bmatrix} \nonumber \\
&= \frac{\gamma}{c} E_p' (\omega') ( b' + i \beta a' )  
\begin{bmatrix}
-\sin \phi' \\ \cos \phi' \\
\end{bmatrix}, \label{eq:34-2} \\
E_{f,z''}'' (\omega') &= -i E_p' (\omega') (a_{out} \sin \theta_-' + a_{in} \sin \theta_+'), \label{eq:34-3} \\
\intertext{and} %
B_{f,z''}'' (\omega') &= 0. \label{eq:34-4}
\end{align}
\end{subequations}
Here, $a'$ and $b'$ in Eq. \eqref{eq:34} are expressed as
\begin{subequations}
\label{eq:35}
\begin{align}
a' &= a_- \cos \theta_-' + a_+ \cos \theta_+', \label{eq:35-1} \\
\intertext{and} %
b' &= b_- \sin \theta_-' + b_+ \sin \theta_+', \label{eq:35-2}
\end{align}
\end{subequations}
with $a_\pm$ and $b_\pm$ defined in Eq. \eqref{eq:22}. Equation \eqref{eq:34} represents the focused electric and magnetic field distributions at a certain angular frequency, but it is still expressed in terms of four-vector components in the boost frame. 

\subsection{\label{sec:level2} Spatio-temporal field distribution in the laboratory frame}
By using the same analogy as in Eqs. \eqref{eq:3} and \eqref{eq:8} and taking the Fourier transformation in the $\omega'$-space, the spatio-temporal field distribution of the RLF in the laboratory frame is obtained as
\begin{subequations}
\label{eq:36}
\begin{align}
\vec{E}_{f}'' &= \int_{-\infty}^{\infty} d \omega' \vec{E}_f'' (\omega') e^{i \omega' t'} = \gamma 
\begin{bmatrix}
( i I_1 + \beta I_2 ) \cos \phi' \\ ( i I_1 + \beta I_2 ) \sin \phi' \\ -i(1/\gamma) I_3 \\
\end{bmatrix}, \label{eq:36-1}
\intertext{and} %
\vec{B}_{f}'' &= \int_{-\infty}^{\infty} d \omega' \vec{B}_f'' (\omega') e^{i \omega' t'} = \frac{\gamma}{c} 
\begin{bmatrix}
-( I_2 + i \beta I_1 ) \sin \phi' \\ ( I_2 + i \beta I_1 ) \cos \phi' \\ 0 \\
\end{bmatrix}.  \label{eq:36-2} 
\end{align}
\end{subequations}
Here, $I_n$ (n=1,2,3) are definite integrals defined as
\begin{subequations}
\label{eq:37}
\begin{align}
I_1 &= \int_{-\infty}^{\infty} d \omega' E_p' (\omega') a' e^{i\omega' t'} \nonumber \\
&= \frac{C_f \sqrt{\mathcal{I}}}{4c} \int_{-\infty}^{\infty} d \omega' \omega' G (\omega') a' e^{i\omega' t'},  \label{eq:37-1} \\ 
I_2 &= \int_{-\infty}^{\infty} d \omega' E_p' (\omega') b' e^{i\omega' t'}  \nonumber \\
&= \frac{C_f \sqrt{\mathcal{I}}}{4c} \int_{-\infty}^{\infty} d \omega' \omega' G (\omega') b' e^{i\omega' t'}, \label{eq:37-2} \\
\intertext{and} %
I_3 &= \frac{C_f \sqrt{\mathcal{I}}}{4c} \int_{-\infty}^{\infty} d \omega' \omega' G (\omega')  (a_{out} \sin \theta_-' + a_{in} \sin \theta_+') e^{i\omega' t'}, \label{eq:37-3}
\end{align}
\end{subequations}
with Eq. \eqref{eq:18}. These integrals can be calculated with the help of Lorentz transformation given by Eq. \eqref{eq:26}. For example, the integral, $I_1$, can be first separated into in-coming and out-going parts as,
\begin{eqnarray} \label{eq:38}
I_1 &&= \frac{C_f \sqrt{\mathcal{I}}}{8 i \rho'} \cos \theta_+' \int_{-\infty}^{\infty} d\omega'  e^{i \omega' \left( t' + \frac{\rho'}{c} \right)} e^{-\frac{(\omega' - \omega_0' )^2}{2 \Delta {\omega'}^2}}  \nonumber \\
&&-\frac{C_f \sqrt{\mathcal{I}}}{8 i \rho'} \cos \theta_-' \int_{-\infty}^{\infty} d\omega'  e^{i \omega' \left( t' - \frac{\rho'}{c} \right)} e^{-\frac{(\omega' - \omega_0' )^2}{2 \Delta {\omega'}^2}}. 
\end{eqnarray}
Then, after applying the Lorentz transformation [Eqs. \eqref{eq:25} and \eqref{eq:30-3}] to the coordinates, we obtain
\begin{eqnarray} \label{eq:39}
I_1 &&= \frac{C_f \sqrt{\mathcal{I}}}{8 i \rho'} \frac{\cos \theta + \beta}{1 + \beta \cos \theta} \int_{-\infty}^{\infty} d\omega' e^{i \frac{\omega' ( t + \rho/c)}{\gamma (1 + \beta \cos \theta)}} e^{-\frac{(\omega' - \omega_0' )^2}{2\Delta {\omega'}^2}} \nonumber \\
&& - \frac{C_f \sqrt{\mathcal{I}}}{8 i \rho'} \frac{\cos \theta - \beta}{1 - \beta \cos \theta}  \int_{-\infty}^{\infty} d\omega' e^{i \frac{\omega' ( t - \rho/c)}{\gamma (1 - \beta \cos \theta)}} e^{-\frac{(\omega' - \omega_0' )^2}{2\Delta {\omega'}^2}}.
\end{eqnarray}
Now, by using the relationship obtained in Eq. \eqref{eq:27} and the linear shift of angular frequency of $\tilde{\omega}_\pm = \omega_\pm'' - \omega_{0,\pm}''$, Eq. \eqref{eq:39} is rewritten in the form of
\begin{eqnarray} \label{eq:40}
I_1 =&& \frac{C_f \sqrt{\mathcal{I}}}{8 i \rho'} \gamma (\cos \theta + \beta)  e^{ i \omega_{0,+}'' \left( t + \frac{\rho}{c} \right) } \nonumber \\
&&\times \int_{-\infty}^{\infty} d \tilde{\omega}_+ \exp \left[ i \tilde{\omega}_+ \left( t + \frac{\rho}{c} \right) \right] \exp \left[ -\frac{\tilde{\omega}_+^2}{\Delta \omega_+''^2} \right] \nonumber \\
&& - \frac{C_f \sqrt{\mathcal{I}}}{8 i \rho'} \gamma (\cos \theta - \beta)  e^{ i \omega_{0,-}'' \left( t - \frac{\rho}{c} \right) } \nonumber \\
&&\times \int_{-\infty}^{\infty} d \tilde{\omega}_- \exp \left[ i \tilde{\omega}_- \left( t - \frac{\rho}{c} \right) \right] \exp \left[ -\frac{\tilde{\omega}_-^2}{\Delta \omega_-''^2} \right] .
\end{eqnarray}
Again, the center frequency and the spectral bandwidth, $\omega_{0,\pm}'$ and $\Delta \omega_\pm''$, in the laboratory frame ($\mathcal{L}_2$) are defined as
\begin{equation} \label{eq:41}
\omega_{0,\pm}'' = \frac{\omega_0'}{\gamma (1 \pm \beta \cos \theta)} \;\mathrm{and}\; \Delta \omega_\pm'' = \frac{\Delta \omega_\pm'}{\gamma (1 \pm \beta \cos \theta)},
\end{equation}
by Eq. \eqref{eq:27}. Using the integral identity \cite{Gradshteyn1} of
\begin{equation} \label{eq:42}
\begin{split}
&\int_0^{\infty} x^{p -1} e^{-q x^2} \cos sx dx  = \\
&\frac{1}{2} q^{- p/2} \Gamma \left( \frac{p}{2} \right) \exp \left( - \frac{s^2}{4q} \right) {_1}F_1 \left(-\frac{p}{2}+\frac{1}{2};\frac{1}{2}; \frac{s^2}{4q} \right),
\end{split}
\end{equation}
the integral, 
\begin{equation} \label{eq:43}
\int_{-\infty}^{\infty} d \tilde{\omega}_\pm \cos \tilde{\omega}_\pm \left( t \pm \frac{\rho}{c} \right) e^ { - \tilde{\omega}_\pm^2 / \Delta {\omega_\pm''}^2} ,
\end{equation}
becomes 
\begin{equation} \label{eq:44}
\sqrt{\pi} \Delta \omega_\pm'' \exp \left[ - \frac{\Delta {\omega_\pm''}^2 (t \pm \rho/c)^2}{4} \right]
\end{equation}
with $p =1$, $q= 1/\Delta {\omega_\pm''}^2$, and $s = t \pm \rho/c$, since $\Gamma(1/2) = \sqrt{\pi}$ and ${_1}F_1 \left(0;1/2; s^2/4q \right) =1$. Here, $\Gamma()$ and ${_1}F_1 ()$ are the Gamma function and the confluent hypergeometric function. Note that $\int_{-\infty}^{\infty} e^{-qx^2} \sin sx dx = 0 $. Finally, after integrating over (-$\infty$,$\infty$)
, Eq. \eqref{eq:40} becomes
\begin{eqnarray} \label{eq:45}
I_1 =&& \sqrt{\pi} \frac{\Delta \omega' C_f}{8 i R} \sqrt{\mathcal{I}} \frac{\cos \theta + \beta}{1 + \beta \cos \theta}  e^{ i \omega_0' \left( T + R/c \right) } \nonumber \\
&& \times \exp \left[ - \frac{\Delta \omega'^2 }{4} \left( T + \frac{R}{c} \right)^2 \right] \nonumber \\
&& - \sqrt{\pi} \frac{\Delta \omega' C_f}{8 i R} \sqrt{\mathcal{I}} \frac{\cos \theta - \beta}{1 - \beta \cos \theta}   e^{ i \omega_0' \left( T - R/c \right) }  \nonumber \\
&& \times \exp \left[ - \frac{\Delta \omega'^2 }{4} \left( T - \frac{R}{c} \right)^2 \right] ,
\end{eqnarray}
with the help of Eq. \eqref{eq:30}. This integral contains information on the in-coming ($T+R/c$) and out-going ($T-R/c$) spherical fields in the $R-T$ space. These in-coming and out-going fields can be expressed with spherical Bessel functions as
\begin{equation} \label{eq:46}
\frac{e^{i \omega_0' (T \pm R/c)}}{R} = \frac{\omega_0'}{c} \left[ y_0 \left( \frac{\omega_0'}{c}R \right) \pm i j_0 \left( \frac{\omega_0'}{c}R \right) \right] e^{i \omega_0' T},
\end{equation}
where $j_0 (\cdot)$ and $y_0 (\cdot)$ are the spherical Bessel functions of the first and the second kinds, respectively. Since the functional value of  $y_0 (\omega_0' R/c)$ is infinity at $R=0$, we take the imaginary part from Eq. \eqref{eq:46} as the solution of Eq. \eqref{eq:45}. Then, we have 
\begin{equation} \label{eq:47}
I_1 =  \frac{\sqrt{\pi} \omega_0' \Delta \omega' C_f \sqrt{\mathcal{I}}}{4c} j_0 \left( \frac{\omega_0'}{c} R \right) \Upsilon_1 e^{i \omega_0' T}.
\end{equation}
with the definition of envelope function of
\begin{equation} \label{eq:48}
\begin{split}
\Upsilon_1 =& \frac{1}{2} \Bigg\{ \frac{\cos \theta + \beta}{1 + \beta \cos \theta} \exp \left[ - \frac{\Delta {\omega'}^2}{4} \left( T + \frac{R}{c} \right)^2 \right]  \\
& + \frac{\cos \theta - \beta}{1 - \beta \cos \theta} \exp \left[ - \frac{\Delta {\omega'}^2}{4} \left( T - \frac{R}{c} \right)^2 \right] \Bigg\}.
\end{split}
\end{equation}	
Similarly, by applying the same mathematical procedures, we obtain the following results for the other integrals as:
\begin{subequations}
\label{eq:49}
\begin{align}
&I_2 = \frac{\sqrt{\pi} \omega_0' \Delta \omega' C_f \sqrt{\mathcal{I}}}{4c} j_1 \left( \frac{\omega_0'}{c} R \right) \Upsilon_2 e^{i \omega_0' T},  \label{eq:49-1} \\ 
\intertext{and} %
&I_3 = \frac{\sqrt{\pi} \omega_0' \Delta \omega' C_f \sqrt{\mathcal{I}}}{4c} j_0 \left( \frac{\omega_0'}{c} R \right) \Upsilon_2 e^{i \omega_0' T}, \label{eq:49-2}
\end{align}
\end{subequations}
with the definition of another envelope function,
\begin{equation} \label{eq:50}
\begin{split}
\Upsilon_2 =& \frac{1}{2} \Bigg\{ \frac{\sin \theta}{\gamma(1 - \beta \cos \theta)} \exp \left[ - \frac{\Delta {\omega'}^2}{4} \left( T - \frac{R}{c} \right)^2 \right]  \\
& + \frac{\sin \theta}{\gamma(1 + \beta \cos \theta)} \exp \left[ - \frac{\Delta {\omega'}^2}{4} \left( T + \frac{R}{c} \right)^2 \right] \Bigg\}.
\end{split}
\end{equation}
Now, inserting Eqs. \eqref{eq:47} and \eqref{eq:49} into Eq. \eqref{eq:36}, the general mathematical expressions for the spatio-temporal field distribution of the RLF with the radial polarization is obtained as,
\begin{subequations}
\label{eq:51}
\begin{align}
\vec{E}_f'' =& \gamma \frac{\sqrt{\pi} \omega_0' \Delta \omega' C_f \sqrt{\mathcal{I}}}{4c}  \nonumber \\
&\times \begin{bmatrix}
\left\{ -j_0 \sin (\omega_0' T) \Upsilon_1 + \beta j_1 \cos (\omega_0' T) \Upsilon_2 \right\} \cos \phi \\ \left\{ -j_0 \sin (\omega_0' T) \Upsilon_1 + \beta j_1 \cos (\omega_0' T) \Upsilon_2 \right\} \sin \phi \\ (1/\gamma) j_0 \sin (\omega_0' T) \Upsilon_2 \\
\end{bmatrix}, \label{eq:51-1} \\
\intertext{and} %
\vec{B}_f'' =& \frac{\gamma}{c} \frac{\sqrt{\pi} \omega_0' \Delta \omega' C_f \sqrt{\mathcal{I}}}{4c}  \nonumber \\
&\times \begin{bmatrix}
-\left\{ j_1 \cos (\omega_0' T) \Upsilon_2 - \beta j_0 \sin (\omega_0' T) \Upsilon_1 \right\} \sin \phi \\ \left\{ j_1 \cos (\omega_0' T) \Upsilon_2 - \beta j_0 \sin (\omega_0' T) \Upsilon_1 \right\} \cos \phi \\ 0 \\
\end{bmatrix}. \label{eq:51-2}
\end{align}
\end{subequations}
in the laboratory frame. In Eq. \eqref{eq:51}, the azimuthal angle, $\phi'$, is replaced by $\phi$ due to $\phi' = \phi$, and $j_{0,1}$ should read $j_{0,1} (\omega' R/c)$. Equation \eqref{eq:51} is valid in the relativistic limit since the $4\pi$-spherical focusing scheme used is valid only when $\gamma \gg 1$.

When the TE mode (azimuthally-polarized) laser pulse is incident and focused by the RFPM, from the symmetry in the polarization, the EM field distributions of the $4\pi$-spherically focused monochromatic TE mode EM wave are expressed as,
\begin{subequations}
\label{eq:52}
\begin{align}
\vec{E}_f ' ({x'}^\mu ; \omega') &= -\hat{\phi'}  E_p ' (\omega ' ) b (\rho', \theta' ; \omega') e ^{i\omega' t'} \nonumber \\
&= \vec{E}_{f,\|}', \label{eq:52-1} \\
\intertext{and} %
\vec{B}_f ' ({x'}^\mu ; \omega') &= \hat{\theta'} i B_p ' (\omega ' ) a (\rho', \theta' ; \omega') e ^{i\omega' t'} \nonumber \\
&= \vec{B}_{f,\perp}' + \vec{B}_{f,\|}'. \label{eq:52-2}
\end{align}
\end{subequations}
In this case, followed by the similar mathematical procedures taken as before, the general mathematical expressions for the spatio-temporal field distribution of the RLF with the azimuthal polarization is obtained as,
\begin{subequations}
\label{eq:53}
\begin{align}
\vec{E}_f'' =& \gamma \frac{\sqrt{\pi} \omega_0' \Delta \omega' C_f \sqrt{\mathcal{I}}}{4c}  \nonumber \\
&\times \begin{bmatrix}
-\left\{ j_1 \cos (\omega_0' T) \Upsilon_2 - \beta j_0 \sin (\omega_0' T) \Upsilon_1 \right\} \sin \phi \\ \left\{ j_1 \cos (\omega_0' T) \Upsilon_2 - \beta j_0 \sin (\omega_0' T) \Upsilon_1 \right\} \cos \phi \\ 0 \\
\end{bmatrix}, \label{eq:53-1} \\
\intertext{and} %
\vec{B}_f'' =& \frac{\gamma}{c} \frac{\sqrt{\pi} \omega_0' \Delta \omega' C_f \sqrt{\mathcal{I}}}{4c}  \nonumber \\
&\times \begin{bmatrix}
\left\{ -j_0 \sin (\omega_0' T) \Upsilon_1 + \beta j_1 \cos (\omega_0' T) \Upsilon_2 \right\} \cos \phi \\ \left\{ -j_0 \sin (\omega_0' T) \Upsilon_1 + \beta j_1 \cos (\omega_0' T) \Upsilon_2 \right\} \sin \phi \\ (1/\gamma) j_0 \sin (\omega_0' T) \Upsilon_2 \\
\end{bmatrix}. \label{eq:53-2}
\end{align}
\end{subequations}
In Eqs. \eqref{eq:51} and \eqref{eq:53}, the peak field strength of the E-field of RLF can be rewritten as
\begin{equation} \label{eq:54}
\begin{split}
\gamma \frac{\sqrt{\pi} \omega_0' \Delta \omega' C_f \sqrt{\mathcal{I}}}{4c} &= \gamma \frac{1+\beta}{1-\beta} \frac{\sqrt{\pi} \omega_0 \Delta \omega C_f \sqrt{\mathcal{I}}}{4c} \\
&= \gamma \frac{1+\beta}{1-\beta} \sqrt{\frac{3 \pi}{c \epsilon_0}} \frac{\pi \omega_0 w_e \sqrt{\mathcal{I}_p}}{4c},
\end{split}
\end{equation}
with the definition of the intensity, $\mathcal{I}_p$ (=$\Delta \omega^2 \mathcal{I}$), in time shown in Sec. 2. So, it is clear that  in the relativistic limit of $\beta \rightarrow 1$ the field strength and the intensity is enhanced by a factor of $\gamma^3 (w_e/\lambda_0)$ and $\gamma^6 (w_e/\lambda_0)^2$ as discovered in \cite{Bulanov3}. Comparing the intensity enhancement given by the RFFM case, the RFPM gives an additional enhancement of a factor of $( 3 \pi^5 /8) \cdot \gamma^2 \cdot (w_e /\lambda_0 )^2 $ with an incident beam size of  $D = 2w_e$. 

The change in the angular frequency of the RLF can be calculated by decomposing the spherical Bessel function into the in-coming and out-going fields again. The phase for the in-coming or out-going field is given by $\omega_0' T \pm \omega_0' R/c$ or $\left[ (1+\beta)/(1 \pm \beta \cos \theta) \right] \omega_0 \left(t \pm \rho/c \right)$ in Eqs. \eqref{eq:51} and \eqref{eq:53}. Thus, the angular frequency for the out-going laser pulse is enhanced by $(1+\beta)/(1-\beta)$ in the forward direction ($\theta = 0$) or $\sim$4$\gamma^2$ in the relativistic limit, which is consistent with the result from the RFFM case. The change in the nominal pulse duration of the out-going laser pulse is determined by the argument of $\Delta {\omega'}^2 \left( T - R/c \right)^2 /4$ in Eq. \eqref{eq:48} or \eqref{eq:50}. From the argument, it is clear that the nominal pulse duration, $\tau_F''$, of the RLF in the laboratory frame is given by $(1/\Delta \omega) (1- \beta \cos \theta)/(1+\beta)$. In the forward direction ($\theta = 0$), the nominal pulse duration is reduced by a factor of $(1-\beta)/(1+\beta)$, which is also consistent with the RFFM case. Although Eqs. \eqref{eq:51} and \eqref{eq:53} well describe the field distribution and its propagation of the RLF, its limitation should be addressed here. In this study, an ideal RFPM, which has a constant velocity and a flat perfect reflectance over the wavelength and incidence angle, is assumed and the recoil effect happening during the reflection of the incident strong laser pulse \cite{Valenta1} is ignored. Therefore, obtaining a mathematical expression for the RLF under a more realistic circumstance will be the next step to be pursued. 

\begin{figure*}[hbt!]
\includegraphics[width=1.8\columnwidth]{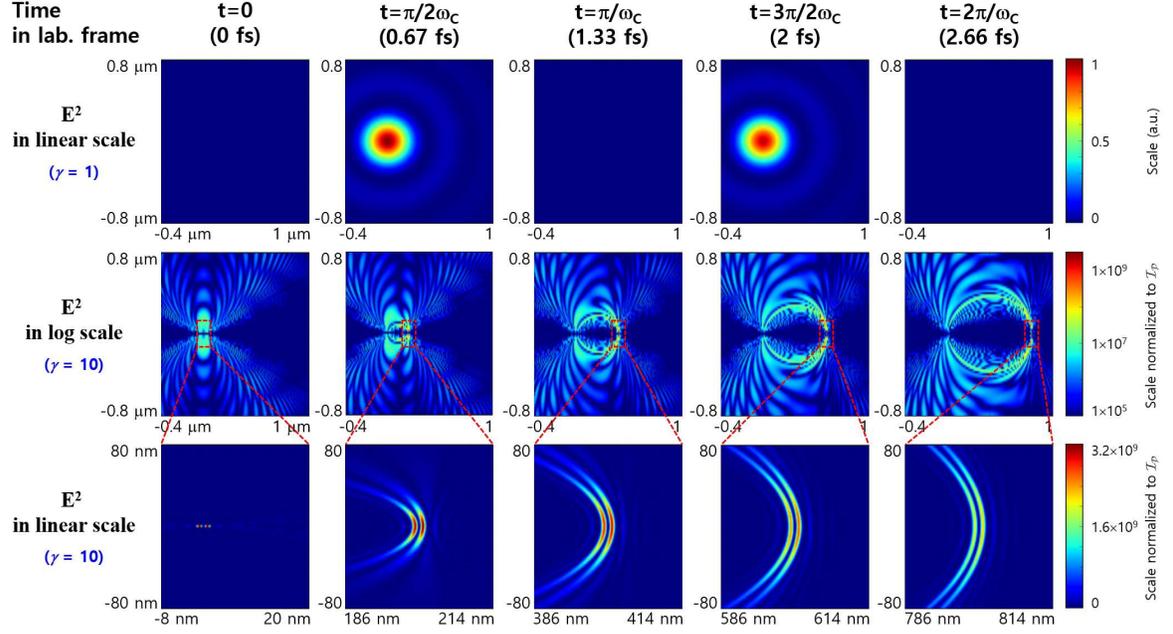}
\caption{\label{fig:wide} The squared electric field distribution calculated from Eq. \eqref{eq:43}. The  squared electric field distribution is expressed in the x (vertical)- z (horizontal) plane. In this plane, $E_{f,y}''$ becomes zero since $\phi = 0$. The electric field distributions in the first row are calculated under $\gamma =1$, i.e., the mirror is stationary. The field distribution agrees well with the characteristics obtained under the $4\pi$-spherical focusing condition. The squared electric field distributions in the second and third rows are calculated under $\gamma =10$. The field distrbution travels with a relativistic velocity of $c\beta = c\sqrt{(\gamma^2 -1)/\gamma^2}$ as shown in the third row, and the peak field strength is enhanced by a factor given by Eq. \eqref{eq:46}.}
\end{figure*}

Figure 3 shows the squared electric field ($E^2 = E_x^2 + E_z^2$) distribution of the RLF at different times. The center frequency ($\omega_0$) of the incident laser pulse is $\sim$2.36$\times$10$^{15}$ rad/s, assuming the center wavelength of 0.8 $\mu$m. The Gaussian width ($\Delta \omega$) of the spectrum is 1.77$\times$10$^{14}$ rad/s, supporting a FWHM pulse duration of $\sim$9.4 fs. The first row in Fig. 3 presents the squared electric field at $\gamma = 1$. The electric and magnetic fields are separated in space and time, and the field oscillates with a period ($\mathcal{T}_{period} = 2 \pi / \omega_0$) of $\sim$2.67 fs. The second row presents the squared electric field at $\gamma = 10$. In this case, the squared field is expressed in the log scale, and it is normalized by the peak laser intensity of the RLF given by the square of Eq. \eqref{eq:54}. The third row presents an enlarged view of the red dashed area in the second row. The spot size of the peak calculated from the second order moment is $\sim$2.5 nm, which is close to the nominal wavelength of 2.0 nm obtained from $\lambda_0/4\gamma^2$. The second and third rows clearly show that the RLF travels at a relativistic speed of $\beta c$ and how its field distribution propagates in time over several hundreds of nm in range. 

\subsection{\label{sec:level2} Recoil effect with a low mirror reflection}
The field calculation was so far based on an ideal mirror which has perfect reflectance, i. e., $\mathcal{R}$=1. However, the reflectance of the mirror is dependent on the mirror model and in general very low \cite{Bulanov4,Pirozhkov1,Kulagin1,Esirkepov2}. Due to a low mirror reflectance, most of the incident (source) pulse energy is transmitted through the mirror, which leads to much lower distortion in the flying mirror than expected. The low reflection of the mirror minimizes the change in mirror shape during reflection. We here explain how the low mirror reflectance reduces the recoil effect on the frequency upshift, and show that the beam radius-wavelength ratio can further intensify the focused intensity toward the nonlinear QED regime even with a low reflectance of the mirror.

From momentum and energy conservation, we re-write Eqs. (3) and (4) of \cite{Valenta1} in two-dimensional form as,
\begin{subequations}
\label{eq:55}
\begin{align}
n_e p_e - n_{\omega} p_{\omega} &= n_e p''_e \cos \theta_e + \mathcal{R}  n_{\omega} p''_{\omega} \cos \theta \nonumber \\ 
& \quad - \left( 1 -  \mathcal{R}  \right) n_{\omega} p_{\omega}, & \label{eq:55-1} \\
0 &= n_e p''_e \sin \theta_e - \mathcal{R} n_{\omega} p''_{\omega} \sin \theta,& \label{eq:55-2} \\
\intertext{where $\theta_e$ and $\theta$ refer to the angles for the electron and photon after reflection, respectively, and} %
n_e \varepsilon_e +n_{\omega} \varepsilon_{\omega} &= n_e \varepsilon''_e + \mathcal{R}  n_{\omega} \varepsilon''_{\omega} + \left( 1 - \mathcal{R} \right) n_{\omega} \varepsilon_{\omega}.& \label{eq:55-3}
\end{align}
\end{subequations}
Here, $p$ and $\varepsilon$ refer to momentum and energy for individual electron and photon, and the subscripts, $e$ and $\omega$, are used to denote electron and photon. $n_e$ and $n_{\omega}$ are population densities for electron and photon. The reflectance, $\mathcal{R}$, depends on the incident angle, but, considering the mathematical simplicity and aperture function, we ignore the angle-dependency for the mirror. The unprimed and double-primed quantities refer to quantities before and after reflection. Eqs. \eqref{eq:55-1} and \eqref{eq:55-2} can be combined by use of $\sin^2 \theta_e + \cos^2 \theta_e = 1$ to yield
\begin{widetext}
\begin{equation} \label{eq:56}
n_e^2 {p''_e}^2 = n_e^2 p_e^2 + \mathcal{R}^2 n_{\omega}^2 {p''_{\omega}}^2 + \mathcal{R}^2 n_{\omega}^2 p_{\omega}^2 -2 \mathcal{R} n_e n_{\omega} p_e p''_{\omega} \cos \theta  - 2 \mathcal{R} n_e n_{\omega} p_e p_{\omega} + 2 \mathcal{R}^2 n_{\omega}^2  p_{\omega} p''_{\omega} \cos \theta,
\end{equation}	
and, subtracting $n_e^2 {p''_e}^2 c^2$ from $n_e^2 {\varepsilon''_e}^2$, we obtain,
\begin{eqnarray} \label{eq:57}
n_e^2 \left( {\varepsilon''_e}^2 - {p''_e}^2 c^2 \right) &=& n_e^2  \left( \varepsilon_e^2 - p_e^2 c^2 \right) + \mathcal{R}^2 n_{\omega}^2  \left( \varepsilon_{\omega}^2 - p_{\omega}^2 c^2 \right) + \mathcal{R}^2 n_{\omega}^2  \left( {\varepsilon''_{\omega}}^2 - {p''_{\omega}}^2 c^2 \right) + 2 \mathcal{R} n_{\omega} n_e \left( \varepsilon_e \varepsilon_{\omega} + p_e p_{\omega} c^2 \right) \nonumber \\
&& - 2 \mathcal{R} n_{\omega} n_e \left( \varepsilon_e \varepsilon''_{\omega} - p_e p''_{\omega} c^2 \cos \theta \right) - 2 \mathcal{R}^2 n_{\omega}^2 \left( \varepsilon_{\omega} \varepsilon''_{\omega} + p_{\omega} p''_{\omega} c^2 \cos \theta \right).
\end{eqnarray}
\end{widetext}
Since ${\varepsilon''_e}^2 - {p''_e}^2 c^2 = \varepsilon_e^2 - p_e^2 c^2 = m_e^2 c^4$ for electrons and ${\varepsilon''_{\omega}}^2 - {p''_{\omega}}^2 c^2 = \varepsilon_{\omega}^2 - p_{\omega}^2 c^2 = 0$ for photons, Eq. \eqref{eq:57} becomes
\begin{equation} \label{eq:58}
\left[ n_e \left( \varepsilon_e - p_e c \cos \theta \right) + \mathcal{R} n_{\omega} \varepsilon_{\omega} \left( 1 + \cos \theta \right) \right] \frac{\varepsilon''_{\omega}}{\varepsilon_{\omega}} = n_e \left( \varepsilon_e + p_e c \right).
\end{equation}
Here, $m_e$ is the electron mass. Then, with the help of $p_e c = \beta \varepsilon_e$, we obtain
\begin{equation} \label{eq:59}
\varepsilon''_{\omega} = \varepsilon_{\omega} \frac{n_e \varepsilon_e (1 + \beta)}{n_e \varepsilon_e (1 - \beta \cos \theta) + \mathcal{R} n_{\omega} \varepsilon_{\omega} (1 + \cos \theta) }.
\end{equation}
The energy density can be further expressed as,
\begin{equation} \label{eq:60}
n_e \varepsilon_e = \gamma n_e m_e c^2, \quad\mathrm{and}\quad  n_{\omega} \varepsilon_{\omega} = \mathcal{I} /c.
\end{equation}
The ratio, $\mathcal{R} n_{\omega} \varepsilon_{\omega} / n_e \varepsilon = \mathcal{R} \mathcal{I} / \gamma n_e m_e c^3$, in Eq. \eqref{eq:59} can be expressed as $\sim 3\times10^{19} \times (\mathcal{R}/n_e)$ at $\mathcal{I}=2.3\times 10^{17}$ W/cm$^2$ in terms of the reflectance and the electron density of the mirror. The reflectance, $\mathcal{R}$, is very low. For example, according to the thin foil electron layer mirror model, the reflectance of the mirror is given by $\mathcal{R}=0.5 \gamma^{-3}$ \cite{Pirozhkov1,Kulagin1,Esirkepov2} and yields 2.75$\times$10$^{-4}$ for a Lorentz $\gamma$ of 12.2. Thus, the frequency for reflected photons can be approximated as,
\begin{eqnarray} \label{eq:61}
\omega'' &\approx& \omega \frac{1+\beta}{1-\beta \cos \theta} \nonumber \\
&&\times \left[ 1  - \frac{\mathcal{R}\mathcal{I}_0}{\gamma n_e m_e c^3} (1 + \cos \theta )^2 e^{- \frac{\sin^2 \theta}{\sin^2 \theta_0} } \right].
\end{eqnarray}
Here, the laser intensity, $\mathcal{I}$, is replaced by $\mathcal{I}_0 \sin^2 \theta e^{-\sin^2 \theta / sin^2 \theta_0}$, which is the incident intensity distribution for the TM or TE mode beam profile. Comparing Eq. \eqref{eq:61} to the well-known frequency upshift formula, $\omega'' = \omega (1+\beta)/(1-\beta \cos \theta) $, the second term in the bracket on the right-hand side gives the correction to the wavelength shift by the recoil effect. Equation \eqref{eq:61} shows how the frequency upshift for the curved mirror is modified by the recoil effect when the mirror reflectance is considered. Since $\mathcal{R} \mathcal{I}_0 / \gamma n_e m_e c^3 \ll 1$ at a source laser power of 180 TW or an intensity, $\mathcal{I}_0$, of 2.3$\times$10$^{17}$ W/cm$^2$, the frequency shift is approximated as $\omega'' \approx \omega (1+\beta)/(1-\beta \cos \theta)$. In addition, the consideration of a low reflectance of $\mathcal{R}$ does not allow violation of the energy balance condition through $n''_{\omega} \hbar \omega'' \approx \mathcal{R} \left( 4 \gamma^2 \mathcal{I}_0/c \right) < \gamma n_e m_e c^2 $. A numerical calculation shows that energy densities [$\mathcal{R} \left( 4 \gamma^2 \mathcal{I}_0/c \right)$ and $\gamma n_e m_e c^2$] for the reflected laser pulse and the electron layer acting as the RFM are $\sim$1.25$\times$10$^6$ J/cm$^3$ and $\sim$1$\times$10$^7$ J/cm$^3$ with a $\gamma$-factor of 12.2, respectively. So, it is valid to apply the approach used in previous subsections when calculating the field distribution, since it does not seriously modify the frequency upshift and the field distribution with a low reflectance. However, the low reflectance affects the reflected energy, consequently the peak intensity of a focused laser field and the $e^+ e^-$ pair production rate as discussed in the following section.

After reflection, the frequency-upshifted source laser pulse is further intensified by the beam radius-wavelength ratio, $w_e / \lambda_0$, as shown in Eqs. \eqref{eq:53} and \eqref{eq:54}. This factor first appeared in the original paper on the RFM \cite{Bulanov4} and results in the intensification of the electromagnetic pulse while maintaining a substantially low source laser intensity on the mirror with a substantially large beam size. With a given reflectance of $\mathcal{R}$, the peak electric field strength of a focused field can be written from Eqs. \eqref{eq:53-1} and \eqref{eq:54} as,
\begin{equation} \label{eq:62}
E''_f = \sqrt{\mathcal{R}} \gamma \frac{1+\beta}{1-\beta} \sqrt{\frac{3 \pi}{c \epsilon_0}} \frac{\pi \omega w_e}{4c} \sqrt{\mathcal{I}}
\end{equation}
Under the condition of $E''_f = E_{Sch} $, Eq. \eqref{eq:62} can be re-expressed as,
\begin{equation} \label{eq:63}
\mathcal{I} = \frac{(\lambda_0 / w_e)^2}{6 \pi^5 R \gamma^6} \mathcal{I}_{Sch}
\end{equation}
with the definition of $\mathcal{I}_{Sch} = (1/2) c \epsilon_0 E_{Sch}^2 \approx 2.3 \times 10^{29}$ W/cm$^2$. Then, the source laser intensity, $\mathcal{I}$, required for reaching the Schwinger field is calculated to be $\sim$2.27$\times$10$^{17}$ W/cm$^2$ with parameters such as $\gamma=12.2$, $\lambda$ = 0.2 $\mu$m, $w_0$ = 156 $\mu$m, and $\mathcal{R} = 0.1\gamma^{-3}$. Thus, the beam radius-wavelength ratio plays a critical role in boosting the focused field strength to the nonlinear QED regime.

\section{Pair production under the RLF field}
Untill now, the analytical field expression for the TM or TE mode RLF has been obtained in the laboratory frame. In this section, the electron-positron ($e^+ e^-$) pair production rate is investigated with the field expressions obtained in the relativistic limit. Here we consider the Schwinger mechanism for the pair production.

\subsection{\label{sec:level2}Invariant fields and pair production rate}
Assuming the Compton wavelength is much less than the wavelength of the RLF \cite{Bulanov1,Lifshitz1,Narozhny1,Aleksandrov1}, the spacetime-dependent $e^+ e^-$ pair production rate, $W_{ep}$, can be calculated from
\begin{equation} \label{eq:64}
W_{ep} = \frac{e^2 E_{Sch}^2}{4 \pi^3 \hbar^2 c} E_{inv} B_{inv} \coth \left( \pi \frac{B_{inv}}{E_{inv}} \right) e^{- \pi/ E_{inv}}.
\end{equation}
Here, $E_{inv}$ and $B_{inv}$ are invariant fields defined by
\begin{subequations}
\label{eq:65}
\begin{align}
E_{inv} = \frac{\sqrt{(\mathcal{F}^2 + \mathcal{G}^2 )^{1/2} - \mathcal{F}}}{E_{Sch}}, \label{eq:65-1} \\
\intertext{and} %
B_{inv} = \frac{\sqrt{(\mathcal{F}^2 + \mathcal{G}^2 )^{1/2} + \mathcal{F}}}{E_{Sch}}. \label{eq:65-2}
\end{align}
\end{subequations}
In Eq. \eqref{eq:65}, the Poincare invariants, $\mathcal{F}$ and $\mathcal{G}$, for the RLF are calculated as 
\begin{equation}
\label{eq:66}
\mathcal{F} = \frac{c^2 B^2 - E^2}{2}, \quad \mathrm{and} \quad \mathcal{G} = c \vec{B} \cdot \vec{E}. 
\end{equation}
Then, the Poincare invariants for the TM mode laser pulse can be calculated with Eq. \eqref{eq:51} and become
\begin{subequations}
\label{eq:67}
\begin{align}
\mathcal{F}_{TM} = &\frac{1}{2} \left( \frac{1+\beta}{1-\beta} \frac{\sqrt{\pi} \omega_0 C_f \sqrt{\mathcal{I}_p}}{4c}  \right)^2 \nonumber \\
&\times \Big[ j_1^2 \cos^2 (\omega_0' T) \Upsilon_2^2 - j_0^2 \sin^2 (\omega_0' T) \left( \Upsilon_1^2 + \Upsilon_2^2 \right) \Big] \nonumber \\
= &\frac{1}{2} \left( \frac{1+\beta}{1-\beta} \frac{\sqrt{\pi} \omega_0 C_f \sqrt{\mathcal{I}_p}}{4c}  \right)^2 j_{\{1-0\}}^2, \label{eq:67-1} \\ 
\intertext{and} %
\mathcal{G}_{TM} &= 0,  \label{eq:67-2}
\end{align}
\end{subequations}
where the function, $j_{\{1-0\}}^2$, is defined as $j_1^2 \cos^2 (\omega_0' T) \Upsilon_2^2 - j_0^2 \sin^2 (\omega_0' T) \left( \Upsilon_1^2 + \Upsilon_2^2 \right)$ and $j_i$, represents the spherical Bessel function, $j_i (\omega_0' \rho'/c)$. The invariant fields, $E_{inv}$ and $H_{inv}$, are determined by the sign of $\mathcal{F}$. When $\mathcal{F} \geq 0$ (i.e., $j_1^2 \cos^2 (\omega_0' T) \Upsilon_2^2 \geq j_0^2 \sin^2 (\omega_0' T) \left( \Upsilon_1^2 + \Upsilon_2^2 \right)$), $\sqrt{\mathcal{F}^2} = \mathcal{F}$. On the other hand, when $\mathcal{F} < 0$, $\sqrt{\mathcal{F}^2} = -\mathcal{F}$. This results in   
\begin{subequations}
\label{eq:68}
\begin{align}
  E_{inv} = &\frac{1+\beta}{1-\beta} \frac{\sqrt{\pi} \omega_0 C_f \sqrt{\mathcal{I}_p}}{4cE_{Sch}}  \begin{cases}
      0 & \text{$\mathcal{F}$ $\geq$ 0}\\
      \sqrt{- j_{\{1-0\}}^2} & \text{$\mathcal{F} <$ 0}\\
    \end{cases} , \label{eq:68-1} \\ 
\intertext{and} %
  B_{inv} = &\frac{1+\beta}{1-\beta} \frac{\sqrt{\pi} \omega_0 C_f \sqrt{\mathcal{I}_p}}{4c^2E_{Sch}}  \begin{cases}
      \sqrt{ j_{\{1-0\}}^2}  & \text{$\mathcal{F}$ $\geq$ 0}\\
      0 & \text{$\mathcal{F} <$ 0}\\
    \end{cases}. \label{eq:68-2}
\end{align}
\end{subequations}
Thus, from Eq. \eqref{eq:64}, it follows that no pair production is expected even with the enhanced field strength of the RLF when $\mathcal{F} \geq 0$. In the other case ($\mathcal{F} < 0$), the $e^+ e^-$ pair production rate via the Schwinger mechanism can be explicitly calculated in terms of the Lorentz $\gamma$-factor, the beam radius-wavelength ratio ($w_e / \lambda_0$), and the laser intensity ($\mathcal{I}_p$) as
\begin{equation} \label{eq:69}
\begin{split}
W_{ep} \approx &12 \pi^2 \alpha \gamma^4 \left( \frac{w_e}{\lambda_0} \right)^2 \left( \frac{\mathcal{I}_p}{\hbar c} \right) \left( -j_{\{1-0\}}^2 \right) \\
&\times \exp \left[ -\frac{1}{\gamma^2} \frac{\lambda_0}{w_e} \frac{ (E_{Sch}/E_p) }{ \sqrt{6 \pi^3} \sqrt{ -j_{\{1-0 \} }^2}} \right], 
\end{split}
\end{equation}
in the relativistic limit. Here, $\alpha$ is the fine structure constant defined as $e^2 / 4 \pi \hbar c \epsilon_0$ and the peak field strength, $E_p$, in time as $\sqrt{2 \mathcal{I}_p / c \epsilon_0}$. The  $e^+ e^-$ pair production occurs only in the region of $j_{\{1-0\}}^2 < 0$. 

For the TE mode laser pulse, the $e^+ e^-$ pair production rate can be calculated with Eq. \eqref{eq:53} and the final form is given by
\begin{equation} \label{eq:70}
\begin{split}
W_{ep} \approx &12 \pi^2 \alpha \gamma^4 \left( \frac{w_e}{\lambda_0} \right)^2 \left( \frac{\mathcal{I}_p}{\hbar c} \right) \left( -j_{\{0-1\}}^2 \right) \\
&\times \exp \left[ -\frac{1}{\gamma^2} \frac{\lambda_0}{w_e} \frac{ (E_{Sch}/E_p) }{ \sqrt{6 \pi^3} \sqrt{ -j_{\{0-1 \} }^2}} \right], 
\end{split}
\end{equation}
where $j_{\{0-1\}}^2$, is defined as $j_0^2 \sin^2 (\omega_0' T) \left( \Upsilon_1^2 + \Upsilon_2^2 \right) - j_1^2 \cos^2 (\omega_0' T) \Upsilon_2^2$. Again, the $e^+ e^-$ pair production occurs only in the region of $j_{\{0-1\}}^2 < 0$. 

Now, let us consider the reflectance, $\mathcal{R}$, of the flying plasma mirror in the calculation of $e^+ e^-$ pair production. The reflectance can be modeled as $0.5 \gamma^{-3}$ for the infinitely thin foil model, $\sim 0.1 \gamma^{-4}$ for the wake wave model, and a rather complicated form can be found for the double-sided mirror model \cite{Kulagin1,Pirozhkov1,Esirkepov2}. By multiplying the reflectance, $\mathcal{R}$, into the intensity, $\mathcal{I}_p$, the pair production rate can be modified as
\begin{subequations}
\label{eq:71}
\begin{align}
W_{ep} \approx &6 \pi^2 \alpha \gamma \left( \frac{w_e}{\lambda_0} \right)^2 \left( \frac{\mathcal{I}_p}{\hbar c} \right) \left( -j_{\{0-1\}}^2 \right) \nonumber \\
& \times \exp \left[ -\frac{1}{\sqrt{\gamma}} \frac{\lambda_0}{w_e} \frac{ (E_{Sch}/E_p) }{ \sqrt{3 \pi^3} \sqrt{ -j_{\{0-1 \} }^2}} \right], \label{eq:71-1} \\ 
\intertext{for the thin foil model or} %
W_{ep} \approx &1.2 \pi^2 \alpha \left( \frac{w_e}{\lambda_0} \right)^2 \left( \frac{\mathcal{I}_p}{\hbar c} \right) \left( -j_{\{0-1\}}^2 \right) \nonumber \\
&\times \exp \left[ -\frac{\lambda_0}{w_e} \frac{ (E_{Sch}/E_p) }{ 0.32 \sqrt{6 \pi^3} \sqrt{ -j_{\{0-1 \} }^2}} \right], \label{eq:71-2}
\end{align}
\end{subequations}
for the wake wave model. From Eq. \eqref{eq:71-2}, it is obvious that, in case of the wake wave model, due to the reflectance of the mirror the pair production rate becomes dependent only on the beam radius-wavelength ratio ($w_e / \lambda_0$). Thus, the $e^+ e^-$ pair production from the thin foil model will be mostly considered. 

The total number, $N_{e^+ e^-}$, of $e^+ e^-$ pairs produced by the RLFs can be estimated by integrating the pair production rate, $W_{ep}$, over a four-volume \cite{Narozhny1}. The threshold of the incident laser power required for a single $e^+ e^-$ pair production is examined with a $\gamma$-factor of 12.2. Such a relativistic mirror can be driven by focusing a 100 PW Ti:S ($\lambda_0$= 0.8 $\mu$m) laser pulse within a focal spot radius of 100 $\mu$m, assuming the $\gamma$-factor of the mirror is given by $\sqrt{1+a_0^2}$. A laser pulse with an $a_0 = 0.1$ and a beam radius, $w_0$, of 100 $\mu m$ is considered as an incident laser pulse to be reflected. According to \cite{Valenta1}, in order to minimize the recoil effect, an incident pulse duration should be less than a characteristic time, $\tau_c$, given as 
\begin{equation} \label{eq:72}
\tau_c = \kappa \frac{2^{4/3} m_e c^2}{2 \Gamma^2 (2/3)} \left( \frac{\omega_0}{\omega_{pe}} \right)^{8/3} \gamma_L^{1/3} \frac{n_e \lambda_L}{\mathcal{I}_p}.
\end{equation}
Here, $\omega_{pe}$ is the Langmuir frequency, $\gamma_L$ the Lorentz $\gamma$-factor for the Langmuir wave, and $\lambda_L$ the Langmuir wavelength. The parameter, $\kappa$, can be obtained from the simulation. Although this analysis is based on an one-dimensional PIC approach, it provides a rough estimation on the maximum pulse duration required for minimizing the recoil effect in time. The calculation with $\kappa  = 1.5\times 10^{-4}$ and an electron density of 0.01 $n_{cr}$ shows that the incident laser pulse with an $a_0$ of 0.1 can have a maximum pulse duration of about 30 fs before the mirror is severely affected by the recoil effect. This pulse duration can be supported by a spectral bandwidth, $\Delta \omega$, of 19 nm ($\sim$32 nm at FWHM). This means that a laser pulse with a pulse duration of $\sim$9.4 fs (supported by a spectral bandwidth, $\Delta \omega$, of 60 nm at 800 nm center wavelength, $\omega_0$) can be used as the incident laser pulse to be reflected. In this case, the minimum reflectance of the mirror required for a single $e^+ e^-$ pairs production is estimated as $\sim$14.1$\%$. The $e^+ e^-$ pairs production is strongly suppressed by the low reflectance (2.7$\times 10^{-2} \%$ for $0.5 \gamma^{-3}$ and 4.5$\times 10^{-4} \%$ for $0.1 \gamma^{-4}$) of the mirror. The required laser field strengths, expressed as $a_0$, increase to $\sim$2.8 for the thin foil and $\sim$21.6 for the wake wave cases. Under these field strengths, the characteristic time given by Eq. \eqref{eq:72} becomes as short as 0.05 fs for the thin foil and to $8 \times 10^{-4}$ fs for the wake wave cases, so the RFPM is destroyed before reaching a field strength required for the single $e^+ e^-$ pair production event.

\subsection{\label{sec:level2}Pair production with two counter-propagating RLFs}
\begin{figure}[b]
\includegraphics[width=1\columnwidth]{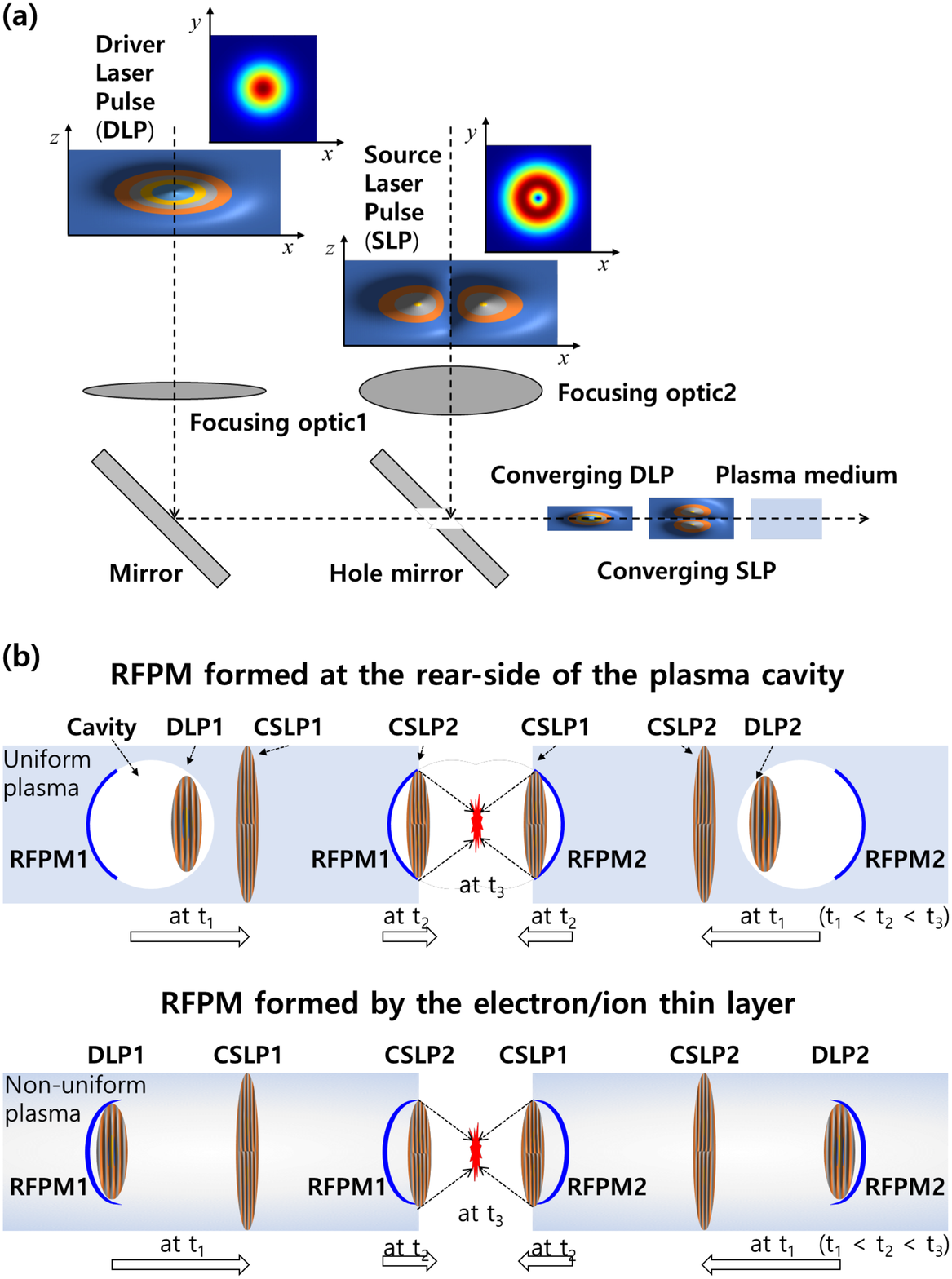}
\caption{\label{fig:epsart} (a) Schematic diagram for collinear coupling of driver and source laser pulses. (b) Formation of RFPMs and collision in vacuum of two source pulses reflected from respective RFPMs. CSLP: Converging source laser pulse.}
\end{figure}
\begin{figure*}
\includegraphics[width=1.8\columnwidth]{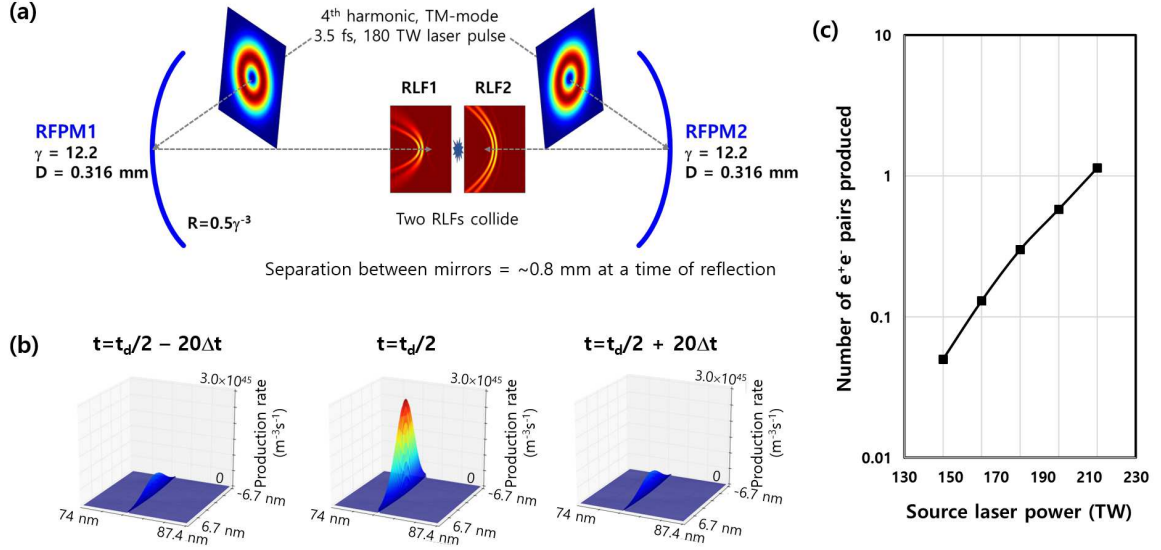}
\caption{\label{fig:wide} (a) The $e^+ e^-$ pair production by colliding two RLFs. The RFPM is driven by a 250 PW laser pulse and the fourth harmonic laser pulse with a center wavelength of 0.2 $\mu$ is incident on the RFPM. The diameter, D, of the mirror is given by 2$w_0$. The two RLFs collide with a proper time delay to maximize the pair production. (b) The calculated $e^+ e^-$ pair production rate, $W_{ep}^T$, when two RLFs collide. A strong enhancement in pair production is observed over the overlapped volume at t = $t_d$/2. (c) The number of $e^+ e^-$ pairs produced as a function of the source laser power.}
\end{figure*}
It is now interesting to consider when two counter-propagating RLFs collide with each other. The experimental set-up considered is visualized in Fig. 4. The two (driver and source) laser pulses are coupled by a hole mirror and propagate in a plasma medium. The source laser pulse (SLP1) proceeds ahead to be reflected by the other counter-propagating RFPM formed by another driver laser pulse (DLP2). Two sets of plasma media are prepared to ensure the collision of reflected SLPs in vacuum. An ideally-sharp edge is preferred. Figure 4(b) shows two different scenarios of forming a RFPM: one is the mirror formed at the rear side of plasma cavity as proposed in \cite{Bulanov3} and the other is the mirror formed by sharp and narrow electron/ion layer driven by the laser pulse \cite{Pirozhkov1,Esirkepov2}. One focus, expressed by   $\vec{E}_f^+ (\rho, t)$ and $\vec{B}_f^+ (\rho, t)$, propagates along the +z axis and the other one, expressed by $\vec{E}_f^- (\rho, t)$ and $\vec{B}_f^- (\rho, t)$, along the -z axis. The + and - symbols in the superscripts are used to express the propagating and the counter-propagating RLFs. TM mode laser pulses are assumed in the calculation, so the total fields are expressed as follows:
\begin{subequations}
\label{eq:73}
\begin{align}
&\vec{E}_f^\pm = \gamma \frac{1+\beta}{1-\beta} \frac{\sqrt{\pi} \omega_0 C_f \sqrt{\mathcal{I}_p}}{4cE_{Sch}} \nonumber \\
&\times \begin{bmatrix}
\left\{ -j_0^\pm \sin (\omega_0' T^\pm) \Upsilon_1^\pm \pm \beta j_1^\pm \cos (\omega_0' T^\pm) \Upsilon_2^\pm \right\} \cos \phi^+ \\ \left\{ -j_0^\pm \sin (\omega_0' T^\pm) \Upsilon_1^\pm \pm \beta j_1^\pm \cos (\omega_0' T^\pm) \Upsilon_2^+ \right\} \sin \phi^\pm \\ (1/\gamma) j_0^\pm \sin (\omega_0' T^\pm) \Upsilon_2^\pm \\
\end{bmatrix}, \label{eq:73-1} \\
\intertext{and} %
&\vec{B}_f^\pm = \frac{\gamma}{c} \frac{1+\beta}{1-\beta} \frac{\sqrt{\pi} \omega_0 C_f \sqrt{\mathcal{I}_p}}{4cE_{Sch}}   \nonumber \\
&\times \begin{bmatrix}
-\left\{\pm j_1^\pm \cos (\omega_0' T^\pm) \Upsilon_2^\pm - \beta j_0^\pm \sin (\omega_0' T^\pm) \Upsilon_1^\pm \right\} \sin \phi^\pm \\ \left\{\pm j_1^\pm \cos (\omega_0' T^\pm) \Upsilon_2^\pm - \beta j_0^\pm \sin (\omega_0' T^\pm) \Upsilon_1^\pm \right\} \cos \phi^\pm \\ 0 \\
\end{bmatrix}. \label{eq:73-2}
\end{align}
\end{subequations}
Here, $j_i^\pm$ = $j_i (\omega_0' R^{\pm}/c)$ and the minus sign in front of $j_1^-$ comes from the change in the sign of magnetic field. The functions, $T^\pm$ and $R^\pm$, for the propagating and counter-propagating RLFs in Eq. \eqref{eq:73} are expressed as, 
\begin{subequations}
\label{eq:74}
\begin{align}
T^\pm (t,\rho) = \frac{t - (\rho/c) \beta \cos \theta^\pm}{\gamma (1-\beta^2 \cos^2 \theta^\pm )}, \label{eq:74-1} \\
\intertext{and} %
R^\pm (\rho,t) = \frac{\rho - ct \beta \cos \theta^\pm}{\gamma (1-\beta^2 \cos^2 \theta^\pm )}. \label{eq:74-2}
\end{align}
\end{subequations}

Since the two RLFs counter-propagate with respect to each other, we find the following relationships for the polar and azimuthal angles between the propagating and the counter-propagating RLFs:
\begin{equation} \label{eq:75}
\theta^- = \pi - \theta^+ = \pi - \theta \quad \mathrm{and}\quad \phi^- = \phi^+ = \phi.
\end{equation}
By using Eq. \eqref{eq:75}, the following relationships for variables between the propagating and counter-propagating RLFs are obtained:
\begin{subequations}
\label{eq:76}
\begin{align}
&T^- (t,\rho) = - T^+ (-t,\rho) \; \mathrm{and}\; R^- (\rho, t) = R^+ (\rho, t), \label{eq:76-1} \\
&T^- (t,\rho) \pm \frac{R^- (\rho, t)}{c} = - \left[ T^+ (-t,\rho) \mp \frac{R^+ (\rho, -t)}{c} \right], \label{eq:76-2} \\
&\frac{\sin \theta^-}{\gamma ( 1 \pm \beta \cos \theta^- )} = \frac{\sin \theta}{\gamma ( 1 \mp \beta \cos \theta) }, \label{eq:76-3} \\
&\frac{\cos \theta^- \pm \beta}{1 \pm \beta \cos \theta^- } = - \frac{\cos \theta \mp \beta}{1 \mp \beta \cos \theta }, \label{eq:76-4} \\
\intertext{and} %
&\Upsilon_1^- (t,\rho) = - \Upsilon_1^+ (-t,\rho) \; \mathrm{and}\; \Upsilon_2^- (t,\rho) =  \Upsilon_2^+ (-t,\rho). \label{eq:76-5}
\end{align}
\end{subequations}
Thus, the total fields, $\vec{E}_f^T$ and $\vec{B}_f^T$, given as the sum of propagating and counter-propagating fields are calculated as,
\begin{subequations}
\label{eq:77}
\begin{align}
&\vec{E}_f^T = \vec{E}_f^+ + \vec{E}_f^- \nonumber \\
&=\gamma \frac{1+\beta}{1-\beta} \frac{\sqrt{\pi} \omega_0 C_f \sqrt{\mathcal{I}_p}}{4cE_{Sch}}   \nonumber \\
&\times \begin{bmatrix}
\left\{ -\left[ \mathcal{J}_0 (t) + \mathcal{J}_0 (-t') \right] + \beta \left[ \mathcal{J}_1 (t) -\mathcal{J}_1 (-t') \right]  \right\} \cos \phi \\ \left\{ -\left[ \mathcal{J}_0 (t) + \mathcal{J}_0 (-t') \right] + \beta \left[ \mathcal{J}_1 (t) -\mathcal{J}_1 (-t') \right]  \right\} \sin \phi \\ (1/\gamma) \left[ \mathcal{J}_0' (t) - \mathcal{J}_0' (-t') \right] \\
\end{bmatrix}, \label{eq:77-1} \\
\intertext{and} %
&\vec{B}_f^T = \vec{B}_f^+ + \vec{B}_f^- \nonumber \\
&=\frac{\gamma}{c} \frac{1+\beta}{1-\beta} \frac{\sqrt{\pi} \omega_0 C_f \sqrt{\mathcal{I}_p}}{4cE_{Sch}} \nonumber \\
&\times \begin{bmatrix}
-\left\{ \left[ \mathcal{J}_1 (t) - \mathcal{J}_1 (-t') \right] - \beta \left[ \mathcal{J}_0 (t) + \mathcal{J}_0 (-t') \right]  \right\} \sin \phi \\ \left\{ \left[ \mathcal{J}_1 (t) - \mathcal{J}_1 (-t') \right] - \beta \left[ \mathcal{J}_0 (t) + \mathcal{J}_0 (-t') \right]  \right\} \cos \phi \\ 0 \\
\end{bmatrix}. \label{eq:77-2}
\end{align}
\end{subequations}
Here, $t'$ is given by $t+t_d$ with a time delay, $t_d$, between two fields. The new functions, $\mathcal{J}_0 (t)$ and $\mathcal{J}_1 (t)$, are defined as,
\begin{subequations}
\label{eq:78}
\begin{align}
&\mathcal{J}_0 (t) = j_0 \left[ \frac{\omega_0' R^+ (\rho,t)}{c} \right] \sin [ \omega_0' T^+ (t,\rho) ] \Upsilon_1^+ (t,\rho), \label{eq:78-1} \\
&\mathcal{J}_0' (t) = j_0 \left[ \frac{\omega_0' R^+ (\rho,t)}{c} \right] \sin [ \omega_0' T^+ (t,\rho) ] \Upsilon_2^+ (t,\rho), \label{eq:78-2} \\
\intertext{and} %
&\mathcal{J}_1 (t) = j_1 \left[ \frac{\omega_0' R^+ (\rho,t)}{c} \right] \cos [ \omega_0' T^+ (t,\rho) ] \Upsilon_2^+ (t,\rho). \label{eq:78-3}
\end{align}
\end{subequations}
Equation \eqref{eq:78} has the same form as Eq. \eqref{eq:51} with the replacement of field components by the ones superposed with two propagating and counter-propagating RLF fields. In this case, the Poincare invariants, $\mathcal{F}^T$ and $\mathcal{G}^T$, are given by
\begin{subequations}
\label{eq:79}
\begin{align}
\mathcal{F}^T =& \frac{1}{2} \left( \frac{1+\beta}{1-\beta} \frac{\sqrt{\pi} \omega_0 C_f \sqrt{\mathcal{I}_p}}{4c}  \right)^2 \nonumber \\
& \times \Big\{ \left[ \mathcal{J}_1 (t) - \mathcal{J}_1 (-t') \right] ^2 - \left[ \mathcal{J}_0 (t) + \mathcal{J}_0 (-t') \right] ^2 \nonumber \\
& -\left[ \mathcal{J}_0' (t) - \mathcal{J}_0' (-t') \right]^2 \Big\} , \label{eq:79-1} \\ 
\intertext{and} %
\mathcal{G}^T &= 0,  \label{eq:79-2}
\end{align}
\end{subequations}
The $e^+ e^-$ pair production rate, $W_{ep}^T$, is calculated as,
\begin{equation} \label{eq:80}
\begin{split}
W_{ep}^T \approx & 12 \pi^2 \alpha \gamma^4 \left( \frac{w_e}{\lambda_0} \right)^2 \left( \frac{\mathcal{I}_p}{\hbar c} \right) \left[ -\mathcal{J}^2 (t,t') \right] \\
&\times \exp \left[ -\frac{1}{\gamma^2} \frac{\lambda_0}{w_e} \frac{ (E_{Sch}/E_p) }{ \sqrt{6 \pi^3} \sqrt{ -\mathcal{J}^2 (t,t')}} \right], 
\end{split}
\end{equation}
when $\mathcal{F}^T < 0$. In Eq. \eqref{eq:80}, the function, $\mathcal{J}^2 (t,t')$, is defined as $\left[ \mathcal{J}_1 (t) - \mathcal{J}_1 (-t') \right] ^2 - \left[ \mathcal{J}_0 (t) + \mathcal{J}_0 (-t') \right] ^2  -\left[ \mathcal{J}_0' (t) - \mathcal{J}_0' (-t') \right]^2$. At t = 0, two RLfs overlap at the origin, and $\mathcal{F}^T$ becomes $-2 \times \left( \frac{1+\beta}{1-\beta} \frac{\sqrt{\pi} \omega_0 C_f \sqrt{\mathcal{I}_p}}{4c}  \right)^2 \mathcal{J}_0^2 (0)$. But, in general, it is not necessarily for two RLFs to be overlapped at t = 0.

The threshold field strength of incident laser pulse required for a single $e^+ e^-$ pair production is examined. In the calculation, two identical RFPMs with a $\gamma$-factor of 12.2 are considered with a mirror reflectance of 2.7$\times 10^{-2} \%$ [see Fig. 5(a)]. Due to the beam radius-wavelength ratio, the high harmonic laser pulse is more favorable than the fundamental wavelength of a high-power laser pulse in reducing the threshold required for the pair production. For the fourth harmonic laser pulse (0.2 $\mu$m) of a 0.8 $\mu$m Ti:S high-power laser pulse. In the numerical calculation of $e^+ e^-$ pairs produced, the $\omega_0' T^\pm (t,\rho)$ and $\omega_0' R^\pm (\rho,t)$ in Eq. \eqref{eq:74} were calculated with $\omega_0' = \sqrt{(1+\beta)/(1-\beta)} \omega_0$. Then, $\mathcal{J}_0 (t)$, $\mathcal{J}_0' (t)$, $\mathcal{J}_1 (t)$, and $\mathcal{J}^2 (t,t')$ in Eqs. \eqref{eq:78} and \eqref{eq:80} were calculated in series with $\omega_0' T^\pm (t,\rho)$ and $\omega_0' R^\pm (\rho,t)$. Next, the $e^+ e^-$ pair production rate, $W_{ep}^T$, in Eq. \eqref{eq:80} was calculated for different time delays. Figure 5(b) shows the calculated $W_{ep}^T$ at different time delays. Finally, the number of $e^+ e^-$ pairs produced for a specific laser [shown in Fig. 5(c)] was obtained after summing up the integrations of  $W_{ep}^T$s over the four-volume obtained for different time delays. The pair production rate is much enhanced in an overlapped volume of two RLFs [see Fig. 5(b)] and the total number of pair produced in the volume dominates. A time that has the maximal pair production rate for a single RLF is chosen as a proper time delay, $t_d$, between two RLFs. A single $e^+ e^-$ pair can be produced at $a_0 = 0.23$ with a $\gamma$-factor of 12.2. The characteristic time, $\tau_c$, for the maximum pulse duration is calculated as about 3.7 fs. This condition can be satisfied by focusing an optical laser pulse (with $\tau_F$ = 3.5 fs and a peak power of 0.18 PW at the fourth harmonic wavelength) within a beam radius of 158 $\mu$m. By assuming the RFPM with a $\gamma$-factor of 12.2 can be driven by focusing a 250 PW laser pulse with a beam radius of 158 $\mu$m, this result implies that electron-positron pair production from two colliding RLFs can be expected with a lower laser power (2$\times$250 PW) laser than that ($\sim$1000 PW level) calculated under the $4\pi$-spherical focusing condition \cite{Gonoskov1,Jeong3}. Again, the determination of characteristic time is based on the one-dimensional model, so the recoil effect in time should be fully understood by a three-dimensional model developed in the near future.

\section{Conclusion}
The mathematical formulae describing the electromagnetic field of the relativistic-flying laser focus formed by an ideal relativistic-flying parabolic mirror were obtained. The main optical characteristics of the relativistic-flying laser focus, such as the enhancement of field strength and its distribution in time and space, could be well understood by the formulae. The field expression of the relativistic-flying laser focus was applied to the estimation of the $e^+ e^-$ pair production by the Schwinger mechanism. The pair production rate under the relativistic-flying laser focus was modified by the Lorentz $\gamma$-factor and the beam radius-wavelength ratio. The calculation shows that even with a strong suppression due to the low reflectance of relativistic-flying parabolic mirror the $e^+ e^-$ pair production is feasible by colliding two counter-propagating relativistic-flying laser focuses at a relatively lower laser power of 250 PW. Although we assume an ideal parabolic shape for the relativistic-flying parabolic mirror, the actual shape of the mirror will have a small deviation from the ideal shape, introducing the wavefront error. The wavefront error deforms the distribution and induces the degradation of the focused intensity. Thus, the next step is to generalize the mathematical formulae for a beam having a wavefront aberration and to calculate the Strehl ratio assessing the focusability. On the other hand, since the formation of the relativistic-flying parabolic mirror will be strongly affected by the instability of the high-power laser pulse and its propagation in the plasma medium, the generation of a stable relativistic-flying parabolic mirror will be a technical challenge for practical applications of the flying mirror. The results obtained can be used to understand the fundamental question on the electron positron pair production from vacuum.

\begin{acknowledgments}
The work was supported by the project High Field Initiative (CZ.02.1.01/0.0/0.0/15\_003/0000449) from the European Regional Development Fund. S. S. Bulanov acknowledges support from the Office of Science of the US DOE under Contract No. DE-AC02-05CH11231. T. Zh. Esirkepov, J. K. Koga, A. S. Pirozhkov, and M. Kando acknowledge support from JSPS KAKENHI Grant Number JP19H00669 and QST President’s Strategic Grant (Creative Research) $\#$20.
\end{acknowledgments}

\nocite{*}

\providecommand{\noopsort}[1]{}\providecommand{\singleletter}[1]{#1}%

\providecommand{\noopsort}[1]{}\providecommand{\singleletter}[1]{#1}%
%


\begin{thebibliography}{62}%
\makeatletter
\providecommand \@ifxundefined [1]{%
 \@ifx{#1\undefined}
}%
\providecommand \@ifnum [1]{%
 \ifnum #1\expandafter \@firstoftwo
 \else \expandafter \@secondoftwo
 \fi
}%
\providecommand \@ifx [1]{%
 \ifx #1\expandafter \@firstoftwo
 \else \expandafter \@secondoftwo
 \fi
}%
\providecommand \natexlab [1]{#1}%
\providecommand \enquote  [1]{``#1''}%
\providecommand \bibnamefont  [1]{#1}%
\providecommand \bibfnamefont [1]{#1}%
\providecommand \citenamefont [1]{#1}%
\providecommand \href@noop [0]{\@secondoftwo}%
\providecommand \href [0]{\begingroup \@sanitize@url \@href}%
\providecommand \@href[1]{\@@startlink{#1}\@@href}%
\providecommand \@@href[1]{\endgroup#1\@@endlink}%
\providecommand \@sanitize@url [0]{\catcode `\\12\catcode `\$12\catcode
  `\&12\catcode `\#12\catcode `\^12\catcode `\_12\catcode `\%12\relax}%
\providecommand \@@startlink[1]{}%
\providecommand \@@endlink[0]{}%
\providecommand \url  [0]{\begingroup\@sanitize@url \@url }%
\providecommand \@url [1]{\endgroup\@href {#1}{\urlprefix }}%
\providecommand \urlprefix  [0]{URL }%
\providecommand \Eprint [0]{\href }%
\providecommand \doibase [0]{https://doi.org/}%
\providecommand \selectlanguage [0]{\@gobble}%
\providecommand \bibinfo  [0]{\@secondoftwo}%
\providecommand \bibfield  [0]{\@secondoftwo}%
\providecommand \translation [1]{[#1]}%
\providecommand \BibitemOpen [0]{}%
\providecommand \bibitemStop [0]{}%
\providecommand \bibitemNoStop [0]{.\EOS\space}%
\providecommand \EOS [0]{\spacefactor3000\relax}%
\providecommand \BibitemShut  [1]{\csname bibitem#1\endcsname}%
\let\auto@bib@innerbib\@empty
\bibitem [{\citenamefont {Strickland}\ and\ \citenamefont
  {Mourou}(1985)}]{Strickland1}%
  \BibitemOpen
  \bibfield  {author} {\bibinfo {author} {\bibfnamefont {D.}~\bibnamefont
  {Strickland}}\ and\ \bibinfo {author} {\bibfnamefont {G.}~\bibnamefont
  {Mourou}},\ }\href@noop {} {\bibfield  {journal} {\bibinfo  {journal} {Opt.
  Commun.}\ }\textbf {\bibinfo {volume} {56}},\ \bibinfo {pages} {219}
  (\bibinfo {year} {1985})}\BibitemShut {NoStop}%
\bibitem [{\citenamefont {Sung}\ \emph {et~al.}(2010)\citenamefont {Sung},
  \citenamefont {Lee}, \citenamefont {Yu}, \citenamefont {Jeong},\ and\
  \citenamefont {Lee}}]{Sung1}%
  \BibitemOpen
  \bibfield  {author} {\bibinfo {author} {\bibfnamefont {J.~H.}\ \bibnamefont
  {Sung}}, \bibinfo {author} {\bibfnamefont {S.~K.}\ \bibnamefont {Lee}},
  \bibinfo {author} {\bibfnamefont {T.~J.}\ \bibnamefont {Yu}}, \bibinfo
  {author} {\bibfnamefont {T.~M.}\ \bibnamefont {Jeong}},\ and\ \bibinfo
  {author} {\bibfnamefont {J.}~\bibnamefont {Lee}},\ }\href@noop {} {\bibfield
  {journal} {\bibinfo  {journal} {Opt. Lett.}\ }\textbf {\bibinfo {volume}
  {35}},\ \bibinfo {pages} {3021} (\bibinfo {year} {2010})}\BibitemShut
  {NoStop}%
\bibitem [{\citenamefont {ELI-beamlines}()}]{ELI}%
  \BibitemOpen
  \bibfield  {author} {\bibinfo {author} {\bibnamefont {ELI-beamlines}},\
  }\href@noop {} {}\bibinfo {howpublished}
  {http://www.extreme-light-infrastructure.eu/}\BibitemShut {NoStop}%
\bibitem [{\citenamefont {Mourou}\ \emph {et~al.}(2006)\citenamefont {Mourou},
  \citenamefont {Tajima},\ and\ \citenamefont {Bulanov}}]{Mourou1}%
  \BibitemOpen
  \bibfield  {author} {\bibinfo {author} {\bibfnamefont {G.~A.}\ \bibnamefont
  {Mourou}}, \bibinfo {author} {\bibfnamefont {T.}~\bibnamefont {Tajima}},\
  and\ \bibinfo {author} {\bibfnamefont {S.~V.}\ \bibnamefont {Bulanov}},\
  }\href@noop {} {\bibfield  {journal} {\bibinfo  {journal} {Rev. Mod. Phys.}\
  }\textbf {\bibinfo {volume} {78}},\ \bibinfo {pages} {309} (\bibinfo {year}
  {2006})}\BibitemShut {NoStop}%
\bibitem [{\citenamefont {Marklund}\ and\ \citenamefont
  {Shukla}(2006)}]{Marklund1}%
  \BibitemOpen
  \bibfield  {author} {\bibinfo {author} {\bibfnamefont {M.}~\bibnamefont
  {Marklund}}\ and\ \bibinfo {author} {\bibfnamefont {P.~K.}\ \bibnamefont
  {Shukla}},\ }\href@noop {} {\bibfield  {journal} {\bibinfo  {journal} {Rev.
  Mod. Phys.}\ }\textbf {\bibinfo {volume} {78}},\ \bibinfo {pages} {591}
  (\bibinfo {year} {2006})}\BibitemShut {NoStop}%
\bibitem [{\citenamefont {Esarey}\ \emph {et~al.}(2009)\citenamefont {Esarey},
  \citenamefont {Schroeder},\ and\ \citenamefont {Leemans}}]{Esarey1}%
  \BibitemOpen
  \bibfield  {author} {\bibinfo {author} {\bibfnamefont {E.}~\bibnamefont
  {Esarey}}, \bibinfo {author} {\bibfnamefont {C.~B.}\ \bibnamefont
  {Schroeder}},\ and\ \bibinfo {author} {\bibfnamefont {W.~P.}\ \bibnamefont
  {Leemans}},\ }\href@noop {} {\bibfield  {journal} {\bibinfo  {journal} {Rev.
  Mod. Phys.}\ }\textbf {\bibinfo {volume} {81}},\ \bibinfo {pages} {1229}
  (\bibinfo {year} {2009})}\BibitemShut {NoStop}%
\bibitem [{\citenamefont {Dunne}(2009)}]{Dunne1}%
  \BibitemOpen
  \bibfield  {author} {\bibinfo {author} {\bibfnamefont {G.~V.}\ \bibnamefont
  {Dunne}},\ }\href@noop {} {\bibfield  {journal} {\bibinfo  {journal} {Eur.
  Phys. J. D}\ }\textbf {\bibinfo {volume} {55}},\ \bibinfo {pages} {327}
  (\bibinfo {year} {2009})}\BibitemShut {NoStop}%
\bibitem [{\citenamefont {DiPiazza}\ \emph {et~al.}(2012)\citenamefont
  {DiPiazza}, \citenamefont {Muller}, \citenamefont {Hatsagortsyan},\ and\
  \citenamefont {Keitel}}]{Piazza1}%
  \BibitemOpen
  \bibfield  {author} {\bibinfo {author} {\bibfnamefont {A.}~\bibnamefont
  {DiPiazza}}, \bibinfo {author} {\bibfnamefont {C.}~\bibnamefont {Muller}},
  \bibinfo {author} {\bibfnamefont {K.~Z.}\ \bibnamefont {Hatsagortsyan}},\
  and\ \bibinfo {author} {\bibfnamefont {C.~H.}\ \bibnamefont {Keitel}},\
  }\href@noop {} {\bibfield  {journal} {\bibinfo  {journal} {Rev. Mod. Phys.}\
  }\textbf {\bibinfo {volume} {84}},\ \bibinfo {pages} {1177} (\bibinfo {year}
  {2012})}\BibitemShut {NoStop}%
\bibitem [{\citenamefont {Blackburn}\ \emph {et~al.}(2018)\citenamefont
  {Blackburn}, \citenamefont {Seipt}, \citenamefont {Bulanov},\ and\
  \citenamefont {Marklund}}]{Blackburn1}%
  \BibitemOpen
  \bibfield  {author} {\bibinfo {author} {\bibfnamefont {T.~G.}\ \bibnamefont
  {Blackburn}}, \bibinfo {author} {\bibfnamefont {D.}~\bibnamefont {Seipt}},
  \bibinfo {author} {\bibfnamefont {S.~S.}\ \bibnamefont {Bulanov}},\ and\
  \bibinfo {author} {\bibfnamefont {M.}~\bibnamefont {Marklund}},\ }\href@noop
  {} {\bibfield  {journal} {\bibinfo  {journal} {Phys. Plasma}\ }\textbf
  {\bibinfo {volume} {25}},\ \bibinfo {pages} {083108} (\bibinfo {year}
  {2018})}\BibitemShut {NoStop}%
\bibitem [{\citenamefont {Klein}\ and\ \citenamefont {Nigam}(1964)}]{Klein1}%
  \BibitemOpen
  \bibfield  {author} {\bibinfo {author} {\bibfnamefont {J.~J.}\ \bibnamefont
  {Klein}}\ and\ \bibinfo {author} {\bibfnamefont {B.~P.}\ \bibnamefont
  {Nigam}},\ }\href@noop {} {\bibfield  {journal} {\bibinfo  {journal} {Phys.
  Rev.}\ }\textbf {\bibinfo {volume} {135}},\ \bibinfo {pages} {B1279}
  (\bibinfo {year} {1964})}\BibitemShut {NoStop}%
\bibitem [{\citenamefont {Karbstein}\ \emph {et~al.}(2015)\citenamefont
  {Karbstein}, \citenamefont {Gies}, \citenamefont {Reuter},\ and\
  \citenamefont {Zepf}}]{Karbstein1}%
  \BibitemOpen
  \bibfield  {author} {\bibinfo {author} {\bibfnamefont {F.}~\bibnamefont
  {Karbstein}}, \bibinfo {author} {\bibfnamefont {H.}~\bibnamefont {Gies}},
  \bibinfo {author} {\bibfnamefont {M.}~\bibnamefont {Reuter}},\ and\ \bibinfo
  {author} {\bibfnamefont {M.}~\bibnamefont {Zepf}},\ }\href@noop {} {\bibfield
   {journal} {\bibinfo  {journal} {Phys. Rev. D}\ }\textbf {\bibinfo {volume}
  {92}},\ \bibinfo {pages} {071301(R)} (\bibinfo {year} {2015})}\BibitemShut
  {NoStop}%
\bibitem [{\citenamefont {Valle}\ \emph {et~al.}(2016)\citenamefont {Valle},
  \citenamefont {Ejlli}, \citenamefont {Gastaldi}, \citenamefont {Messineo},
  \citenamefont {Milotti}, \citenamefont {Pengo}, \citenamefont {Ruoso},\ and\
  \citenamefont {Zavattini}}]{Valle1}%
  \BibitemOpen
  \bibfield  {author} {\bibinfo {author} {\bibfnamefont {F.~D.}\ \bibnamefont
  {Valle}}, \bibinfo {author} {\bibfnamefont {A.}~\bibnamefont {Ejlli}},
  \bibinfo {author} {\bibfnamefont {U.}~\bibnamefont {Gastaldi}}, \bibinfo
  {author} {\bibfnamefont {G.}~\bibnamefont {Messineo}}, \bibinfo {author}
  {\bibfnamefont {E.}~\bibnamefont {Milotti}}, \bibinfo {author} {\bibfnamefont
  {R.}~\bibnamefont {Pengo}}, \bibinfo {author} {\bibfnamefont
  {G.}~\bibnamefont {Ruoso}},\ and\ \bibinfo {author} {\bibfnamefont
  {G.}~\bibnamefont {Zavattini}},\ }\href@noop {} {\bibfield  {journal}
  {\bibinfo  {journal} {Eur. Phys. J. C}\ }\textbf {\bibinfo {volume} {76}},\
  \bibinfo {pages} {24} (\bibinfo {year} {2016})}\BibitemShut {NoStop}%
\bibitem [{\citenamefont {Shen}\ \emph {et~al.}(2018)\citenamefont {Shen},
  \citenamefont {Bu}, \citenamefont {Xu}, \citenamefont {Xu}, \citenamefont
  {Ji}, \citenamefont {Li},\ and\ \citenamefont {Xu}}]{Shen1}%
  \BibitemOpen
  \bibfield  {author} {\bibinfo {author} {\bibfnamefont {B.}~\bibnamefont
  {Shen}}, \bibinfo {author} {\bibfnamefont {Z.}~\bibnamefont {Bu}}, \bibinfo
  {author} {\bibfnamefont {J.}~\bibnamefont {Xu}}, \bibinfo {author}
  {\bibfnamefont {T.}~\bibnamefont {Xu}}, \bibinfo {author} {\bibfnamefont
  {L.}~\bibnamefont {Ji}}, \bibinfo {author} {\bibfnamefont {R.}~\bibnamefont
  {Li}},\ and\ \bibinfo {author} {\bibfnamefont {Z.}~\bibnamefont {Xu}},\
  }\href@noop {} {\bibfield  {journal} {\bibinfo  {journal} {Plasma Phys.
  Control. Fusion}\ }\textbf {\bibinfo {volume} {60}},\ \bibinfo {pages}
  {044002} (\bibinfo {year} {2018})}\BibitemShut {NoStop}%
\bibitem [{\citenamefont {Karplus}\ and\ \citenamefont
  {Neuman}(1950)}]{Karplus1}%
  \BibitemOpen
  \bibfield  {author} {\bibinfo {author} {\bibfnamefont {R.}~\bibnamefont
  {Karplus}}\ and\ \bibinfo {author} {\bibfnamefont {M.}~\bibnamefont
  {Neuman}},\ }\href@noop {} {\bibfield  {journal} {\bibinfo  {journal} {Phys.
  Rev.}\ }\textbf {\bibinfo {volume} {80}},\ \bibinfo {pages} {380} (\bibinfo
  {year} {1950})}\BibitemShut {NoStop}%
\bibitem [{\citenamefont {Karplus}\ and\ \citenamefont
  {Neuman}(1951)}]{Karplus2}%
  \BibitemOpen
  \bibfield  {author} {\bibinfo {author} {\bibfnamefont {R.}~\bibnamefont
  {Karplus}}\ and\ \bibinfo {author} {\bibfnamefont {M.}~\bibnamefont
  {Neuman}},\ }\href@noop {} {\bibfield  {journal} {\bibinfo  {journal} {Phys.
  Rev.}\ }\textbf {\bibinfo {volume} {83}},\ \bibinfo {pages} {776} (\bibinfo
  {year} {1951})}\BibitemShut {NoStop}%
\bibitem [{\citenamefont {Tollis}(1964)}]{Tollis1}%
  \BibitemOpen
  \bibfield  {author} {\bibinfo {author} {\bibfnamefont {B.~D.}\ \bibnamefont
  {Tollis}},\ }\href@noop {} {\bibfield  {journal} {\bibinfo  {journal} {Nuovo
  Cim.}\ }\textbf {\bibinfo {volume} {32}},\ \bibinfo {pages} {757} (\bibinfo
  {year} {1964})}\BibitemShut {NoStop}%
\bibitem [{\citenamefont {Tollis}(1965)}]{Tollis2}%
  \BibitemOpen
  \bibfield  {author} {\bibinfo {author} {\bibfnamefont {B.~D.}\ \bibnamefont
  {Tollis}},\ }\href@noop {} {\bibfield  {journal} {\bibinfo  {journal} {Nuovo
  Cim.}\ }\textbf {\bibinfo {volume} {35}},\ \bibinfo {pages} {1182} (\bibinfo
  {year} {1965})}\BibitemShut {NoStop}%
\bibitem [{\citenamefont {Lundstr{\"o}m}\ \emph {et~al.}(2006)\citenamefont
  {Lundstr{\"o}m}, \citenamefont {Brodin}, \citenamefont {Lundin},
  \citenamefont {Marklund}, \citenamefont {Bingham}, \citenamefont {Collier},
  \citenamefont {Mendonca},\ and\ \citenamefont {Norreys}}]{Lundstrom1}%
  \BibitemOpen
  \bibfield  {author} {\bibinfo {author} {\bibfnamefont {E.}~\bibnamefont
  {Lundstr{\"o}m}}, \bibinfo {author} {\bibfnamefont {G.}~\bibnamefont
  {Brodin}}, \bibinfo {author} {\bibfnamefont {J.}~\bibnamefont {Lundin}},
  \bibinfo {author} {\bibfnamefont {M.}~\bibnamefont {Marklund}}, \bibinfo
  {author} {\bibfnamefont {R.}~\bibnamefont {Bingham}}, \bibinfo {author}
  {\bibfnamefont {J.}~\bibnamefont {Collier}}, \bibinfo {author} {\bibfnamefont
  {J.~T.}\ \bibnamefont {Mendonca}},\ and\ \bibinfo {author} {\bibfnamefont
  {P.}~\bibnamefont {Norreys}},\ }\href@noop {} {\bibfield  {journal} {\bibinfo
   {journal} {Phys. Rev. Lett.}\ }\textbf {\bibinfo {volume} {96}},\ \bibinfo
  {pages} {083602} (\bibinfo {year} {2006})}\BibitemShut {NoStop}%
\bibitem [{\citenamefont {Bulanov}\ \emph {et~al.}(2011)\citenamefont
  {Bulanov}, \citenamefont {Esirkepov}, \citenamefont {Hayashi}, \citenamefont
  {Kando}, \citenamefont {Kiriyama}, \citenamefont {Koga}, \citenamefont
  {Kondo}, \citenamefont {Kotaki}, \citenamefont {Pirozhkov}, \citenamefont
  {Bulanov}, \citenamefont {Zhidkov}, \citenamefont {Chen}, \citenamefont
  {Neely}, \citenamefont {Kato}, \citenamefont {Narozhny},\ and\ \citenamefont
  {Korn}}]{Bulanov1}%
  \BibitemOpen
  \bibfield  {author} {\bibinfo {author} {\bibfnamefont {S.~V.}\ \bibnamefont
  {Bulanov}}, \bibinfo {author} {\bibfnamefont {T.~Zh.}\ \bibnamefont
  {Esirkepov}}, \bibinfo {author} {\bibfnamefont {Y.}~\bibnamefont {Hayashi}},
  \bibinfo {author} {\bibfnamefont {M.}~\bibnamefont {Kando}}, \bibinfo
  {author} {\bibfnamefont {H.}~\bibnamefont {Kiriyama}}, \bibinfo {author}
  {\bibfnamefont {J.~K.}\ \bibnamefont {Koga}}, \bibinfo {author}
  {\bibfnamefont {K.}~\bibnamefont {Kondo}}, \bibinfo {author} {\bibfnamefont
  {H.}~\bibnamefont {Kotaki}}, \bibinfo {author} {\bibfnamefont {A.~S.}\
  \bibnamefont {Pirozhkov}}, \bibinfo {author} {\bibfnamefont {S.~S.}\
  \bibnamefont {Bulanov}}, \bibinfo {author} {\bibfnamefont {A.~G.}\
  \bibnamefont {Zhidkov}}, \bibinfo {author} {\bibfnamefont {P.}~\bibnamefont
  {Chen}}, \bibinfo {author} {\bibfnamefont {D.}~\bibnamefont {Neely}},
  \bibinfo {author} {\bibfnamefont {Y.}~\bibnamefont {Kato}}, \bibinfo {author}
  {\bibfnamefont {N.~B.}\ \bibnamefont {Narozhny}},\ and\ \bibinfo {author}
  {\bibfnamefont {G.}~\bibnamefont {Korn}},\ }\href@noop {} {\bibfield
  {journal} {\bibinfo  {journal} {Nucl. Instrum. Methods Phys. Res. A}\
  }\textbf {\bibinfo {volume} {660}},\ \bibinfo {pages} {31} (\bibinfo {year}
  {2011})}\BibitemShut {NoStop}%
\bibitem [{\citenamefont {Koga}\ \emph {et~al.}(2012)\citenamefont {Koga},
  \citenamefont {Bulanov}, \citenamefont {Esirkepov}, \citenamefont
  {Pirozhkov}, \citenamefont {Kando},\ and\ \citenamefont {Rosanov}}]{Koga1}%
  \BibitemOpen
  \bibfield  {author} {\bibinfo {author} {\bibfnamefont {J.~K.}~\bibnamefont
  {Koga}}, \bibinfo {author} {\bibfnamefont {S.~V.}~\bibnamefont {Bulanov}},
  \bibinfo {author} {\bibfnamefont {T.~Zh.}\ \bibnamefont {Esirkepov}}, \bibinfo
  {author} {\bibfnamefont {A.~S.}~\bibnamefont {Pirozhkov}}, \bibinfo {author}
  {\bibfnamefont {M.}~\bibnamefont {Kando}},\ and\ \bibinfo {author}
  {\bibfnamefont {N.~N.}~\bibnamefont {Rosanov}},\ }\href@noop {} {\bibfield
  {journal} {\bibinfo  {journal} {Phys. Rev. A}\ }\textbf {\bibinfo {volume}
  {86}},\ \bibinfo {pages} {053823} (\bibinfo {year} {2012})}\BibitemShut
  {NoStop}%
\bibitem [{\citenamefont {Jeong}\ \emph
  {et~al.}(2020{\natexlab{a}})\citenamefont {Jeong}, \citenamefont {Bulanov},
  \citenamefont {Sasorov}, \citenamefont {Korn}, \citenamefont {Koga},\ and\
  \citenamefont {Bulanov}}]{Jeong1}%
  \BibitemOpen
  \bibfield  {author} {\bibinfo {author} {\bibfnamefont {T.~M.}\ \bibnamefont
  {Jeong}}, \bibinfo {author} {\bibfnamefont {S.~V.}\ \bibnamefont {Bulanov}},
  \bibinfo {author} {\bibfnamefont {P.~V.}\ \bibnamefont {Sasorov}}, \bibinfo
  {author} {\bibfnamefont {G.}~\bibnamefont {Korn}}, \bibinfo {author}
  {\bibfnamefont {J.~K.}\ \bibnamefont {Koga}},\ and\ \bibinfo {author}
  {\bibfnamefont {S.~S.}\ \bibnamefont {Bulanov}},\ }\href@noop {} {\bibfield
  {journal} {\bibinfo  {journal} {Phys. Rev. A}\ }\textbf {\bibinfo {volume}
  {102}},\ \bibinfo {pages} {023504} (\bibinfo {year}
  {2020}{\natexlab{a}})}\BibitemShut {NoStop}%
\bibitem [{\citenamefont {Schwinger}(1951)}]{Schwinger1}%
  \BibitemOpen
  \bibfield  {author} {\bibinfo {author} {\bibfnamefont {J.}~\bibnamefont
  {Schwinger}},\ }\href@noop {} {\bibfield  {journal} {\bibinfo  {journal}
  {Phys. Rev.}\ }\textbf {\bibinfo {volume} {82}},\ \bibinfo {pages} {664}
  (\bibinfo {year} {1951})}\BibitemShut {NoStop}%
\bibitem [{\citenamefont {Bulanov}\ \emph {et~al.}(2010)\citenamefont
  {Bulanov}, \citenamefont {Mur}, \citenamefont {Narozhny}, \citenamefont
  {Nees},\ and\ \citenamefont {Popov}}]{SSBulanov1}%
  \BibitemOpen
  \bibfield  {author} {\bibinfo {author} {\bibfnamefont {S.~S.}\ \bibnamefont
  {Bulanov}}, \bibinfo {author} {\bibfnamefont {V.~D.}\ \bibnamefont {Mur}},
  \bibinfo {author} {\bibfnamefont {N.~B.}\ \bibnamefont {Narozhny}}, \bibinfo
  {author} {\bibfnamefont {J.}~\bibnamefont {Nees}},\ and\ \bibinfo {author}
  {\bibfnamefont {V.~S.}\ \bibnamefont {Popov}},\ }\href@noop {} {\bibfield
  {journal} {\bibinfo  {journal} {Phys. Rev. Lett.}\ }\textbf {\bibinfo
  {volume} {104}},\ \bibinfo {pages} {220404} (\bibinfo {year}
  {2010})}\BibitemShut {NoStop}%
\bibitem [{\citenamefont {Gonoskov}\ \emph {et~al.}(2013)\citenamefont
  {Gonoskov}, \citenamefont {Gonoskov}, \citenamefont {Harvey}, \citenamefont
  {Ilderton}, \citenamefont {Kim}, \citenamefont {Marklund}, \citenamefont
  {Mourou},\ and\ \citenamefont {Sergeev}}]{Gonoskov1}%
  \BibitemOpen
  \bibfield  {author} {\bibinfo {author} {\bibfnamefont {A.}~\bibnamefont
  {Gonoskov}}, \bibinfo {author} {\bibfnamefont {I.}~\bibnamefont {Gonoskov}},
  \bibinfo {author} {\bibfnamefont {C.}~\bibnamefont {Harvey}}, \bibinfo
  {author} {\bibfnamefont {A.}~\bibnamefont {Ilderton}}, \bibinfo {author}
  {\bibfnamefont {A.}~\bibnamefont {Kim}}, \bibinfo {author} {\bibfnamefont
  {M.}~\bibnamefont {Marklund}}, \bibinfo {author} {\bibfnamefont
  {G.}~\bibnamefont {Mourou}},\ and\ \bibinfo {author} {\bibfnamefont
  {A.}~\bibnamefont {Sergeev}},\ }\href@noop {} {\bibfield  {journal} {\bibinfo
   {journal} {Phys. Rev. Lett.}\ }\textbf {\bibinfo {volume} {111}},\ \bibinfo
  {pages} {060404} (\bibinfo {year} {2013})}\BibitemShut {NoStop}%
\bibitem [{\citenamefont {Yu}\ \emph {et~al.}(2019)\citenamefont {Yu},
  \citenamefont {Lu}, \citenamefont {Takahashi}, \citenamefont {Hu},
  \citenamefont {Gong}, \citenamefont {Ma}, \citenamefont {Huang},
  \citenamefont {Chen},\ and\ \citenamefont {Yan}}]{Yu1}%
  \BibitemOpen
  \bibfield  {author} {\bibinfo {author} {\bibfnamefont {J.~Q.}\ \bibnamefont
  {Yu}}, \bibinfo {author} {\bibfnamefont {H.~Y.}\ \bibnamefont {Lu}}, \bibinfo
  {author} {\bibfnamefont {T.}~\bibnamefont {Takahashi}}, \bibinfo {author}
  {\bibfnamefont {R.~H.}\ \bibnamefont {Hu}}, \bibinfo {author} {\bibfnamefont
  {Z.}~\bibnamefont {Gong}}, \bibinfo {author} {\bibfnamefont {W.~J.}\
  \bibnamefont {Ma}}, \bibinfo {author} {\bibfnamefont {Y.~S.}\ \bibnamefont
  {Huang}}, \bibinfo {author} {\bibfnamefont {C.~E.}\ \bibnamefont {Chen}},\
  and\ \bibinfo {author} {\bibfnamefont {X.~Q.}\ \bibnamefont {Yan}},\
  }\href@noop {} {\bibfield  {journal} {\bibinfo  {journal} {Phys. Rev. Lett.}\
  }\textbf {\bibinfo {volume} {122}},\ \bibinfo {pages} {014802} (\bibinfo
  {year} {2019})}\BibitemShut {NoStop}%
\bibitem [{\citenamefont {SULF}()}]{SULF}%
  \BibitemOpen
  \bibfield  {author} {\bibinfo {author} {\bibnamefont {SULF}},\ }\href@noop {}
  {}\bibinfo {howpublished} {http://english.siom.cas.cn/}\BibitemShut {NoStop}%
\bibitem [{\citenamefont {XCELS}()}]{XCELS}%
  \BibitemOpen
  \bibfield  {author} {\bibinfo {author} {\bibnamefont {XCELS}},\ }\href@noop
  {} {}\bibinfo {howpublished} {http://www.xcels.iapras.ru/}\BibitemShut
  {NoStop}%
\bibitem [{\citenamefont {Danson}\ \emph {et~al.}(2019)\citenamefont {Danson},
  \citenamefont {Haefner}, \citenamefont {Bromage}, \citenamefont {Butcher},
  \citenamefont {Chanteloup}, \citenamefont {Chowdhury}, \citenamefont
  {Galvanauskas}, \citenamefont {Gizzi}, \citenamefont {Hein}, \citenamefont
  {Hillier},\ and\ \citenamefont {et~al.}}]{Danson1}%
  \BibitemOpen
  \bibfield  {author} {\bibinfo {author} {\bibfnamefont {C.~N.}\ \bibnamefont
  {Danson}}, \bibinfo {author} {\bibfnamefont {C.}~\bibnamefont {Haefner}},
  \bibinfo {author} {\bibfnamefont {J.}~\bibnamefont {Bromage}}, \bibinfo
  {author} {\bibfnamefont {T.}~\bibnamefont {Butcher}}, \bibinfo {author}
  {\bibfnamefont {J.-C.~F.}\ \bibnamefont {Chanteloup}}, \bibinfo {author}
  {\bibfnamefont {E.~A.}\ \bibnamefont {Chowdhury}}, \bibinfo {author}
  {\bibfnamefont {A.}~\bibnamefont {Galvanauskas}}, \bibinfo {author}
  {\bibfnamefont {L.~A.}\ \bibnamefont {Gizzi}}, \bibinfo {author}
  {\bibfnamefont {J.}~\bibnamefont {Hein}}, \bibinfo {author} {\bibfnamefont
  {D.~I.}\ \bibnamefont {Hillier}},\ and\ \bibinfo {author} {\bibnamefont
  {et~al.}},\ }\href@noop {} {\bibfield  {journal} {\bibinfo  {journal} {High
  Power Laser Sci. Eng.}\ }\textbf {\bibinfo {volume} {7}},\ \bibinfo {pages}
  {e54} (\bibinfo {year} {2019})}\BibitemShut {NoStop}%
\bibitem [{\citenamefont {Li}\ \emph {et~al.}(2021)\citenamefont {Li},
  \citenamefont {Kato},\ and\ \citenamefont {Kawanaka}}]{Li1}%
  \BibitemOpen
  \bibfield  {author} {\bibinfo {author} {\bibfnamefont {Z.}~\bibnamefont
  {Li}}, \bibinfo {author} {\bibfnamefont {Y.}~\bibnamefont {Kato}},\ and\
  \bibinfo {author} {\bibfnamefont {J.}~\bibnamefont {Kawanaka}},\ }\href@noop
  {} {\bibfield  {journal} {\bibinfo  {journal} {Sci. Rep.}\ }\textbf {\bibinfo
  {volume} {11}},\ \bibinfo {pages} {151} (\bibinfo {year} {2021})}\BibitemShut
  {NoStop}%
\bibitem [{\citenamefont {Bahk}\ \emph {et~al.}(2004)\citenamefont {Bahk},
  \citenamefont {Rousseau}, \citenamefont {Planchon}, \citenamefont {Chvykov},
  \citenamefont {Kalintchenko}, \citenamefont {Maksimchuk}, \citenamefont
  {Mourou},\ and\ \citenamefont {Yanovsky}}]{Bahk1}%
  \BibitemOpen
  \bibfield  {author} {\bibinfo {author} {\bibfnamefont {S.}~\bibnamefont
  {Bahk}}, \bibinfo {author} {\bibfnamefont {P.}~\bibnamefont {Rousseau}},
  \bibinfo {author} {\bibfnamefont {T.}~\bibnamefont {Planchon}}, \bibinfo
  {author} {\bibfnamefont {V.}~\bibnamefont {Chvykov}}, \bibinfo {author}
  {\bibfnamefont {G.}~\bibnamefont {Kalintchenko}}, \bibinfo {author}
  {\bibfnamefont {A.}~\bibnamefont {Maksimchuk}}, \bibinfo {author}
  {\bibfnamefont {G.}~\bibnamefont {Mourou}},\ and\ \bibinfo {author}
  {\bibfnamefont {V.}~\bibnamefont {Yanovsky}},\ }\href@noop {} {\bibfield
  {journal} {\bibinfo  {journal} {Opt. Lett.}\ }\textbf {\bibinfo {volume}
  {29}},\ \bibinfo {pages} {2837} (\bibinfo {year} {2004})}\BibitemShut
  {NoStop}%
\bibitem [{\citenamefont {Jeong}\ \emph {et~al.}(2015)\citenamefont {Jeong},
  \citenamefont {Weber}, \citenamefont {Garrec}, \citenamefont {Margarone},
  \citenamefont {Mocek},\ and\ \citenamefont {Korn}}]{Jeong2}%
  \BibitemOpen
  \bibfield  {author} {\bibinfo {author} {\bibfnamefont {T.~M.}\ \bibnamefont
  {Jeong}}, \bibinfo {author} {\bibfnamefont {S.}~\bibnamefont {Weber}},
  \bibinfo {author} {\bibfnamefont {B.~L.}\ \bibnamefont {Garrec}}, \bibinfo
  {author} {\bibfnamefont {D.}~\bibnamefont {Margarone}}, \bibinfo {author}
  {\bibfnamefont {T.}~\bibnamefont {Mocek}},\ and\ \bibinfo {author}
  {\bibfnamefont {G.}~\bibnamefont {Korn}},\ }\href@noop {} {\bibfield
  {journal} {\bibinfo  {journal} {Opt. Express}\ }\textbf {\bibinfo {volume}
  {23}},\ \bibinfo {pages} {11641} (\bibinfo {year} {2015})}\BibitemShut
  {NoStop}%
\bibitem [{\citenamefont {Gonoskov}\ \emph {et~al.}(2012)\citenamefont
  {Gonoskov}, \citenamefont {Aiello}, \citenamefont {Heugel},\ and\
  \citenamefont {Leuchs}}]{Gonoskov2}%
  \BibitemOpen
  \bibfield  {author} {\bibinfo {author} {\bibfnamefont {I.}~\bibnamefont
  {Gonoskov}}, \bibinfo {author} {\bibfnamefont {A.}~\bibnamefont {Aiello}},
  \bibinfo {author} {\bibfnamefont {S.}~\bibnamefont {Heugel}},\ and\ \bibinfo
  {author} {\bibfnamefont {G.}~\bibnamefont {Leuchs}},\ }\href@noop {}
  {\bibfield  {journal} {\bibinfo  {journal} {Phys. Rev. A}\ }\textbf {\bibinfo
  {volume} {86}},\ \bibinfo {pages} {053836} (\bibinfo {year}
  {2012})}\BibitemShut {NoStop}%
\bibitem [{\citenamefont {Jeong}\ \emph
  {et~al.}(2020{\natexlab{b}})\citenamefont {Jeong}, \citenamefont {Bulanov},
  \citenamefont {Sasorov}, \citenamefont {Bulanov}, \citenamefont {Koga},\ and\
  \citenamefont {Korn}}]{Jeong3}%
  \BibitemOpen
  \bibfield  {author} {\bibinfo {author} {\bibfnamefont {T.~M.}\ \bibnamefont
  {Jeong}}, \bibinfo {author} {\bibfnamefont {S.~V.}\ \bibnamefont {Bulanov}},
  \bibinfo {author} {\bibfnamefont {P.}~\bibnamefont {Sasorov}}, \bibinfo
  {author} {\bibfnamefont {S.~S.}\ \bibnamefont {Bulanov}}, \bibinfo {author}
  {\bibfnamefont {J.~K.}\ \bibnamefont {Koga}},\ and\ \bibinfo {author}
  {\bibfnamefont {G.}~\bibnamefont {Korn}},\ }\href@noop {} {\bibfield
  {journal} {\bibinfo  {journal} {Opt. Express}\ }\textbf {\bibinfo {volume}
  {28}},\ \bibinfo {pages} {13991} (\bibinfo {year}
  {2020}{\natexlab{b}})}\BibitemShut {NoStop}%
\bibitem [{\citenamefont {Ritus}(1985)}]{Ritus1}%
  \BibitemOpen
  \bibfield  {author} {\bibinfo {author} {\bibfnamefont {V.~I.}\ \bibnamefont
  {Ritus}},\ }\href@noop {} {\bibfield  {journal} {\bibinfo  {journal} {J. Sov.
  Laser Res.}\ }\textbf {\bibinfo {volume} {6}},\ \bibinfo {pages} {497}
  (\bibinfo {year} {1985})}\BibitemShut {NoStop}%
\bibitem [{\citenamefont {Burke}\ \emph {et~al.}(1997)\citenamefont {Burke},
  \citenamefont {Field}, \citenamefont {Horton-Smith}, \citenamefont {Spencer},
  \citenamefont {Walz}, \citenamefont {Berridge}, \citenamefont {Bugg},
  \citenamefont {Shmakov}, \citenamefont {Weidemann}, \citenamefont {Bula},
  \citenamefont {McDonald}, \citenamefont {Prebys}, \citenamefont {Bamber},
  \citenamefont {Boege}, \citenamefont {Koffas}, \citenamefont {Kotseroglou},
  \citenamefont {Melissinos}, \citenamefont {Meyerhofer}, \citenamefont
  {Reis},\ and\ \citenamefont {Raggk}}]{Burke1}%
  \BibitemOpen
  \bibfield  {author} {\bibinfo {author} {\bibfnamefont {D.~L.}\ \bibnamefont
  {Burke}}, \bibinfo {author} {\bibfnamefont {R.~C.}\ \bibnamefont {Field}},
  \bibinfo {author} {\bibfnamefont {G.}~\bibnamefont {Horton-Smith}}, \bibinfo
  {author} {\bibfnamefont {J.~E.}\ \bibnamefont {Spencer}}, \bibinfo {author}
  {\bibfnamefont {D.}~\bibnamefont {Walz}}, \bibinfo {author} {\bibfnamefont
  {S.~C.}\ \bibnamefont {Berridge}}, \bibinfo {author} {\bibfnamefont {W.~M.}\
  \bibnamefont {Bugg}}, \bibinfo {author} {\bibfnamefont {K.}~\bibnamefont
  {Shmakov}}, \bibinfo {author} {\bibfnamefont {A.~W.}\ \bibnamefont
  {Weidemann}}, \bibinfo {author} {\bibfnamefont {C.}~\bibnamefont {Bula}},
  \bibinfo {author} {\bibfnamefont {K.~T.}\ \bibnamefont {McDonald}}, \bibinfo
  {author} {\bibfnamefont {E.~J.}\ \bibnamefont {Prebys}}, \bibinfo {author}
  {\bibfnamefont {C.}~\bibnamefont {Bamber}}, \bibinfo {author} {\bibfnamefont
  {S.~J.}\ \bibnamefont {Boege}}, \bibinfo {author} {\bibfnamefont
  {T.}~\bibnamefont {Koffas}}, \bibinfo {author} {\bibfnamefont
  {T.}~\bibnamefont {Kotseroglou}}, \bibinfo {author} {\bibfnamefont {A.~C.}\
  \bibnamefont {Melissinos}}, \bibinfo {author} {\bibfnamefont {D.~D.}\
  \bibnamefont {Meyerhofer}}, \bibinfo {author} {\bibfnamefont {D.~A.}\
  \bibnamefont {Reis}},\ and\ \bibinfo {author} {\bibfnamefont
  {W.}~\bibnamefont {Ragg}},\ }\href@noop {} {\bibfield  {journal} {\bibinfo
  {journal} {Phys. Rev. Lett.}\ }\textbf {\bibinfo {volume} {79}},\ \bibinfo
  {pages} {1626} (\bibinfo {year} {1997})}\BibitemShut {NoStop}%
\bibitem [{\citenamefont {Marklund}\ and\ \citenamefont
  {Lundin}(2009)}]{Marklund2}%
  \BibitemOpen
  \bibfield  {author} {\bibinfo {author} {\bibfnamefont {M.}~\bibnamefont
  {Marklund}}\ and\ \bibinfo {author} {\bibfnamefont {J.}~\bibnamefont
  {Lundin}},\ }\href@noop {} {\bibfield  {journal} {\bibinfo  {journal} {Eur.
  Phys. J. D}\ }\textbf {\bibinfo {volume} {55}},\ \bibinfo {pages} {319}
  (\bibinfo {year} {2009})}\BibitemShut {NoStop}%
\bibitem [{\citenamefont {King}\ \emph {et~al.}(2018)\citenamefont {King},
  \citenamefont {Hu},\ and\ \citenamefont {Shen}}]{King1}%
  \BibitemOpen
  \bibfield  {author} {\bibinfo {author} {\bibfnamefont {B.}~\bibnamefont
  {King}}, \bibinfo {author} {\bibfnamefont {H.}~\bibnamefont {Hu}},\ and\
  \bibinfo {author} {\bibfnamefont {B.}~\bibnamefont {Shen}},\ }\href@noop {}
  {\bibfield  {journal} {\bibinfo  {journal} {Phys. Rev. A}\ }\textbf {\bibinfo
  {volume} {98}},\ \bibinfo {pages} {023817} (\bibinfo {year}
  {2018})}\BibitemShut {NoStop}%
\bibitem [{\citenamefont {Baumann}\ \emph {et~al.}(2019)\citenamefont
  {Baumann}, \citenamefont {Nerush}, \citenamefont {Pukhov},\ and\
  \citenamefont {Kostyukov}}]{Baumann1}%
  \BibitemOpen
  \bibfield  {author} {\bibinfo {author} {\bibfnamefont {C.}~\bibnamefont
  {Baumann}}, \bibinfo {author} {\bibfnamefont {E.~N.}\ \bibnamefont {Nerush}},
  \bibinfo {author} {\bibfnamefont {A.}~\bibnamefont {Pukhov}},\ and\ \bibinfo
  {author} {\bibfnamefont {I.~Y.}\ \bibnamefont {Kostyukov}},\ }\href@noop {}
  {\bibfield  {journal} {\bibinfo  {journal} {Sci. Rep.}\ }\textbf {\bibinfo
  {volume} {9}},\ \bibinfo {pages} {9407} (\bibinfo {year} {2019})}\BibitemShut
  {NoStop}%
\bibitem [{\citenamefont {Piazza}\ \emph {et~al.}(2019)\citenamefont {Piazza},
  \citenamefont {Tamburini}, \citenamefont {Meuren},\ and\ \citenamefont
  {Keitel}}]{Piazza2}%
  \BibitemOpen
  \bibfield  {author} {\bibinfo {author} {\bibfnamefont {A.}\ \bibnamefont
  {DiPiazza}}, \bibinfo {author} {\bibfnamefont {M.}~\bibnamefont {Tamburini}},
  \bibinfo {author} {\bibfnamefont {S.}~\bibnamefont {Meuren}},\ and\ \bibinfo
  {author} {\bibfnamefont {C.~H.}\ \bibnamefont {Keitel}},\ }\href@noop {}
  {\bibfield  {journal} {\bibinfo  {journal} {Phys. Rev. A}\ }\textbf {\bibinfo
  {volume} {99}},\ \bibinfo {pages} {022125} (\bibinfo {year}
  {2019})}\BibitemShut {NoStop}%
\bibitem [{\citenamefont {Bulanov}\ \emph {et~al.}(2013)\citenamefont
  {Bulanov}, \citenamefont {Esirkepov}, \citenamefont {Kando}, \citenamefont
  {Pirozhkov},\ and\ \citenamefont {Rosanov}}]{Bulanov2}%
  \BibitemOpen
  \bibfield  {author} {\bibinfo {author} {\bibfnamefont {S.~V.}\ \bibnamefont
  {Bulanov}}, \bibinfo {author} {\bibfnamefont {T.~Zh.}\ \bibnamefont
  {Esirkepov}}, \bibinfo {author} {\bibfnamefont {M.}~\bibnamefont {Kando}},
  \bibinfo {author} {\bibfnamefont {A.~S.}\ \bibnamefont {Pirozhkov}},\ and\
  \bibinfo {author} {\bibfnamefont {N.~N.}\ \bibnamefont {Rosanov}},\
  }\href@noop {} {\bibfield  {journal} {\bibinfo  {journal} {Physics Uspekhi}\
  }\textbf {\bibinfo {volume} {56}},\ \bibinfo {pages} {429} (\bibinfo {year}
  {2013})}\BibitemShut {NoStop}%
\bibitem [{\citenamefont {Quere}\ and\ \citenamefont
  {Vincenti}(2021)}]{Quere1}%
  \BibitemOpen
  \bibfield  {author} {\bibinfo {author} {\bibfnamefont {F.}~\bibnamefont
  {Quere}}\ and\ \bibinfo {author} {\bibfnamefont {H.}~\bibnamefont
  {Vincenti}},\ }\href@noop {} {\bibfield  {journal} {\bibinfo  {journal} {High
  Power Laser Sci. Eng.}\ }\textbf {\bibinfo {volume} {9}},\ \bibinfo {pages}
  {e6} (\bibinfo {year} {2021})}\BibitemShut {NoStop}%
\bibitem [{\citenamefont {Einstein}(1905)}]{Einstein1}%
  \BibitemOpen
  \bibfield  {author} {\bibinfo {author} {\bibfnamefont {A.}~\bibnamefont
  {Einstein}},\ }\href@noop {} {\bibfield  {journal} {\bibinfo  {journal} {Ann.
  Phys.}\ }\textbf {\bibinfo {volume} {17}},\ \bibinfo {pages} {891} (\bibinfo
  {year} {1905})}\BibitemShut {NoStop}%
\bibitem [{\citenamefont {Bulanov}\ \emph {et~al.}(2003)\citenamefont
  {Bulanov}, \citenamefont {Esirkepov},\ and\ \citenamefont
  {Tajima}}]{Bulanov3}%
  \BibitemOpen
  \bibfield  {author} {\bibinfo {author} {\bibfnamefont {S.~V.}\ \bibnamefont
  {Bulanov}}, \bibinfo {author} {\bibfnamefont {T.~Zh.}\ \bibnamefont {Esirkepov}},\
  and\ \bibinfo {author} {\bibfnamefont {T.}~\bibnamefont {Tajima}},\
  }\href@noop {} {\bibfield  {journal} {\bibinfo  {journal} {Phys. Rev. Lett.}\
  }\textbf {\bibinfo {volume} {91}},\ \bibinfo {pages} {085001} (\bibinfo
  {year} {2003})}\BibitemShut {NoStop}%
\bibitem [{\citenamefont {Zhang}\ \emph {et~al.}(2020)\citenamefont {Zhang},
  \citenamefont {Bulanov}, \citenamefont {Seipt}, \citenamefont {Arefiev},\
  and\ \citenamefont {Thomas}}]{Zhang1}%
  \BibitemOpen
  \bibfield  {author} {\bibinfo {author} {\bibfnamefont {P.}~\bibnamefont
  {Zhang}}, \bibinfo {author} {\bibfnamefont {S.~S.}\ \bibnamefont {Bulanov}},
  \bibinfo {author} {\bibfnamefont {D.}~\bibnamefont {Seipt}}, \bibinfo
  {author} {\bibfnamefont {A.~V.}\ \bibnamefont {Arefiev}},\ and\ \bibinfo
  {author} {\bibfnamefont {A.~G.~R.}\ \bibnamefont {Thomas}},\ }\href@noop {}
  {\bibfield  {journal} {\bibinfo  {journal} {Phys. Plasma}\ }\textbf {\bibinfo
  {volume} {27}},\ \bibinfo {pages} {050601} (\bibinfo {year}
  {2020})}\BibitemShut {NoStop}%
\bibitem [{\citenamefont {Kando}\ \emph {et~al.}(2009)\citenamefont {Kando},
  \citenamefont {Pirozhkov}, \citenamefont {Kawase}, \citenamefont {Esirkepov},
  \citenamefont {Fukuda}, \citenamefont {Kiriyama}, \citenamefont {Okada},
  \citenamefont {Daito}, \citenamefont {Kameshima}, \citenamefont {Hayashi},
  \citenamefont {Kotaki}, \citenamefont {Mori}, \citenamefont {Koga},
  \citenamefont {Daido}, \citenamefont {Faenov}, \citenamefont {Pikuz},
  \citenamefont {Ma}, \citenamefont {Chen}, \citenamefont {Ragozin},
  \citenamefont {Kawachi}, \citenamefont {Kato}, \citenamefont {Tajima},\ and\
  \citenamefont {Bulanov}}]{Kando1}%
  \BibitemOpen
  \bibfield  {author} {\bibinfo {author} {\bibfnamefont {M.}~\bibnamefont
  {Kando}}, \bibinfo {author} {\bibfnamefont {A.~S.}\ \bibnamefont
  {Pirozhkov}}, \bibinfo {author} {\bibfnamefont {K.}~\bibnamefont {Kawase}},
  \bibinfo {author} {\bibfnamefont {T.~Zh.}\ \bibnamefont {Esirkepov}}, \bibinfo
  {author} {\bibfnamefont {Y.}~\bibnamefont {Fukuda}}, \bibinfo {author}
  {\bibfnamefont {H.}~\bibnamefont {Kiriyama}}, \bibinfo {author}
  {\bibfnamefont {H.}~\bibnamefont {Okada}}, \bibinfo {author} {\bibfnamefont
  {I.}~\bibnamefont {Daito}}, \bibinfo {author} {\bibfnamefont
  {T.}~\bibnamefont {Kameshima}}, \bibinfo {author} {\bibfnamefont
  {Y.}~\bibnamefont {Hayashi}}, \bibinfo {author} {\bibfnamefont
  {H.}~\bibnamefont {Kotaki}}, \bibinfo {author} {\bibfnamefont
  {M.}~\bibnamefont {Mori}}, \bibinfo {author} {\bibfnamefont {J.~K.}\
  \bibnamefont {Koga}}, \bibinfo {author} {\bibfnamefont {H.}~\bibnamefont
  {Daido}}, \bibinfo {author} {\bibfnamefont {A.~Y.}\ \bibnamefont {Faenov}},
  \bibinfo {author} {\bibfnamefont {T.}~\bibnamefont {Pikuz}}, \bibinfo
  {author} {\bibfnamefont {J.}~\bibnamefont {Ma}}, \bibinfo {author}
  {\bibfnamefont {L.-M.}\ \bibnamefont {Chen}}, \bibinfo {author}
  {\bibfnamefont {E.~N.}\ \bibnamefont {Ragozin}}, \bibinfo {author}
  {\bibfnamefont {T.}~\bibnamefont {Kawachi}}, \bibinfo {author} {\bibfnamefont
  {Y.}~\bibnamefont {Kato}}, \bibinfo {author} {\bibfnamefont {T.}~\bibnamefont
  {Tajima}},\ and\ \bibinfo {author} {\bibfnamefont {S.~V.}\ \bibnamefont
  {Bulanov}},\ }\href@noop {} {\bibfield  {journal} {\bibinfo  {journal} {Phys.
  Rev. Lett.}\ }\textbf {\bibinfo {volume} {103}},\ \bibinfo {pages} {235003}
  (\bibinfo {year} {2009})}\BibitemShut {NoStop}%
\bibitem [{\citenamefont {Bulanov}\ \emph {et~al.}(2016)\citenamefont
  {Bulanov}, \citenamefont {Esirkepov}, \citenamefont {Kando},\ and\
  \citenamefont {Koga}}]{Bulanov4}%
  \BibitemOpen
  \bibfield  {author} {\bibinfo {author} {\bibfnamefont {S.~V.}\ \bibnamefont
  {Bulanov}}, \bibinfo {author} {\bibfnamefont {T.~Zh.}\ \bibnamefont
  {Esirkepov}}, \bibinfo {author} {\bibfnamefont {M.}~\bibnamefont {Kando}},\
  and\ \bibinfo {author} {\bibfnamefont {J.}~\bibnamefont {Koga}},\ }\href@noop
  {} {\bibfield  {journal} {\bibinfo  {journal} {Plasma Sources Sci. Technol.}\
  }\textbf {\bibinfo {volume} {25}},\ \bibinfo {pages} {053001} (\bibinfo
  {year} {2016})}\BibitemShut {NoStop}%
\bibitem [{\citenamefont {Koga}\ \emph {et~al.}(2018)\citenamefont {Koga},
  \citenamefont {Bulanov}, \citenamefont {Esirkepov}, \citenamefont {Kando},
  \citenamefont {Bulanov},\ and\ \citenamefont {Pirozhkov}}]{Koga2}%
  \BibitemOpen
  \bibfield  {author} {\bibinfo {author} {\bibfnamefont {J.}~\bibnamefont
  {Koga}}, \bibinfo {author} {\bibfnamefont {S.~V.}\ \bibnamefont {Bulanov}},
  \bibinfo {author} {\bibfnamefont {T.~Zh.}\ \bibnamefont {Esirkepov}}, \bibinfo
  {author} {\bibfnamefont {M.}~\bibnamefont {Kando}}, \bibinfo {author}
  {\bibfnamefont {S.~S.}\ \bibnamefont {Bulanov}},\ and\ \bibinfo {author}
  {\bibfnamefont {A.}~\bibnamefont {Pirozhkov}},\ }\href@noop {} {\bibfield
  {journal} {\bibinfo  {journal} {Plasma Phys. Control. Fusion}\ }\textbf
  {\bibinfo {volume} {60}},\ \bibinfo {pages} {074007} (\bibinfo {year}
  {2018})}\BibitemShut {NoStop}%
\bibitem [{\citenamefont {Esirkepov}\ \emph {et~al.}(2020)\citenamefont
  {Esirkepov}, \citenamefont {Mu}, \citenamefont {Gu}, \citenamefont {Jeong},
  \citenamefont {Valenta}, \citenamefont {Klimo}, \citenamefont {Koga},
  \citenamefont {Kando}, \citenamefont {Neely}, \citenamefont {Korn},
  \citenamefont {Bulanov},\ and\ \citenamefont {Pirozhkov}}]{Esirkepov1}%
  \BibitemOpen
  \bibfield  {author} {\bibinfo {author} {\bibfnamefont {T.~Zh.}\ \bibnamefont
  {Esirkepov}}, \bibinfo {author} {\bibfnamefont {J.}~\bibnamefont {Mu}},
  \bibinfo {author} {\bibfnamefont {Y.}~\bibnamefont {Gu}}, \bibinfo {author}
  {\bibfnamefont {T.~M.}\ \bibnamefont {Jeong}}, \bibinfo {author}
  {\bibfnamefont {P.}~\bibnamefont {Valenta}}, \bibinfo {author} {\bibfnamefont
  {O.}~\bibnamefont {Klimo}}, \bibinfo {author} {\bibfnamefont {J.~K.}\
  \bibnamefont {Koga}}, \bibinfo {author} {\bibfnamefont {M.}~\bibnamefont
  {Kando}}, \bibinfo {author} {\bibfnamefont {D.}~\bibnamefont {Neely}},
  \bibinfo {author} {\bibfnamefont {G.}~\bibnamefont {Korn}}, \bibinfo {author}
  {\bibfnamefont {S.~V.}\ \bibnamefont {Bulanov}},\ and\ \bibinfo {author}
  {\bibfnamefont {A.~S.}\ \bibnamefont {Pirozhkov}},\ }\href@noop {} {\bibfield
   {journal} {\bibinfo  {journal} {Phys. Plasmas}\ }\textbf {\bibinfo {volume}
  {27}},\ \bibinfo {pages} {052103} (\bibinfo {year} {2020})}\BibitemShut
  {NoStop}%
\bibitem [{\citenamefont {Mu}\ \emph {et~al.}(2020)\citenamefont {Mu},
  \citenamefont {Esirkepov}, \citenamefont {Valenta}, \citenamefont {Gu},
  \citenamefont {Jeong}, \citenamefont {Pirozhkov}, \citenamefont {Koga},
  \citenamefont {Kando}, \citenamefont {Korn},\ and\ \citenamefont
  {Bulanov}}]{Mu1}%
  \BibitemOpen
  \bibfield  {author} {\bibinfo {author} {\bibfnamefont {J.}~\bibnamefont
  {Mu}}, \bibinfo {author} {\bibfnamefont {T.~Zh.}\ \bibnamefont {Esirkepov}},
  \bibinfo {author} {\bibfnamefont {P.}~\bibnamefont {Valenta}}, \bibinfo
  {author} {\bibfnamefont {Y.}~\bibnamefont {Gu}}, \bibinfo {author}
  {\bibfnamefont {T.~M.}\ \bibnamefont {Jeong}}, \bibinfo {author}
  {\bibfnamefont {A.~S.}\ \bibnamefont {Pirozhkov}}, \bibinfo {author}
  {\bibfnamefont {J.~K.}\ \bibnamefont {Koga}}, \bibinfo {author}
  {\bibfnamefont {M.}~\bibnamefont {Kando}}, \bibinfo {author} {\bibfnamefont
  {G.}~\bibnamefont {Korn}},\ and\ \bibinfo {author} {\bibfnamefont {S.~V.}\
  \bibnamefont {Bulanov}},\ }\href@noop {} {\bibfield  {journal} {\bibinfo
  {journal} {Phys. Rev. E}\ }\textbf {\bibinfo {volume} {102}},\ \bibinfo
  {pages} {053202} (\bibinfo {year} {2020})}\BibitemShut {NoStop}%
\bibitem [{\citenamefont {Froula}\ \emph {et~al.}(2018)\citenamefont {Froula},
  \citenamefont {Turnbull}, \citenamefont {Davies},\ and\ \citenamefont
  {et~al.}}]{Froula1}%
  \BibitemOpen
  \bibfield  {author} {\bibinfo {author} {\bibfnamefont {D.~H.}\ \bibnamefont
  {Froula}}, \bibinfo {author} {\bibfnamefont {D.}~\bibnamefont {Turnbull}},
  \bibinfo {author} {\bibfnamefont {A.~S.}\ \bibnamefont {Davies}},\ and\
  \bibinfo {author} {\bibnamefont {et~al.}},\ }\href@noop {} {\bibfield
  {journal} {\bibinfo  {journal} {Nature Photon.}\ }\textbf {\bibinfo {volume}
  {12}},\ \bibinfo {pages} {262} (\bibinfo {year} {2018})}\BibitemShut
  {NoStop}%
\bibitem [{\citenamefont {Jeong}\ \emph {et~al.}(2018)\citenamefont {Jeong},
  \citenamefont {Bulanov}, \citenamefont {Weber},\ and\ \citenamefont
  {Korn}}]{Jeong4}%
  \BibitemOpen
  \bibfield  {author} {\bibinfo {author} {\bibfnamefont {T.~M.}\ \bibnamefont
  {Jeong}}, \bibinfo {author} {\bibfnamefont {S.~V.}\ \bibnamefont {Bulanov}},
  \bibinfo {author} {\bibfnamefont {S.}~\bibnamefont {Weber}},\ and\ \bibinfo
  {author} {\bibfnamefont {G.}~\bibnamefont {Korn}},\ }\href@noop {} {\bibfield
   {journal} {\bibinfo  {journal} {Opt. Express}\ }\textbf {\bibinfo {volume}
  {26}},\ \bibinfo {pages} {33091} (\bibinfo {year} {2018})}\BibitemShut
  {NoStop}%
\bibitem [{\citenamefont {Dunne}\ \emph {et~al.}(2009)\citenamefont {Dunne},
  \citenamefont {Gies},\ and\ \citenamefont {Schutzhold}}]{Dunne2}%
  \BibitemOpen
  \bibfield  {author} {\bibinfo {author} {\bibfnamefont {G.~V.}\ \bibnamefont
  {Dunne}}, \bibinfo {author} {\bibfnamefont {H.}~\bibnamefont {Gies}},\ and\
  \bibinfo {author} {\bibfnamefont {R.}~\bibnamefont {Schutzhold}},\
  }\href@noop {} {\bibfield  {journal} {\bibinfo  {journal} {Phys. Rev. D}\
  }\textbf {\bibinfo {volume} {80}},\ \bibinfo {pages} {111301(R)} (\bibinfo
  {year} {2009})}\BibitemShut {NoStop}%
\bibitem [{\citenamefont {Pirozhkov}\ \emph {et~al.}(2007)\citenamefont
  {Pirozhkov}, \citenamefont {Ma}, \citenamefont {Kando}, \citenamefont
  {Esirkepov}, \citenamefont {Fukuda}, \citenamefont {Chen}, \citenamefont
  {Daito}, \citenamefont {Ogura}, \citenamefont {Homma}, \citenamefont
  {Hayashi}, \citenamefont {H.Kotaki}, \citenamefont {Sagisaka}, \citenamefont
  {Mori}, \citenamefont {Koga}, \citenamefont {Kawachi}, \citenamefont {Daido},
  \citenamefont {Bulanov}, \citenamefont {Kimura}, \citenamefont {Kato},\ and\
  \citenamefont {Tajima}}]{Pirozhkov1}%
  \BibitemOpen
  \bibfield  {author} {\bibinfo {author} {\bibfnamefont {A.~S.}\ \bibnamefont
  {Pirozhkov}}, \bibinfo {author} {\bibfnamefont {J.}~\bibnamefont {Ma}},
  \bibinfo {author} {\bibfnamefont {M.}~\bibnamefont {Kando}}, \bibinfo
  {author} {\bibfnamefont {T.~Zh.}\ \bibnamefont {Esirkepov}}, \bibinfo {author}
  {\bibfnamefont {Y.}~\bibnamefont {Fukuda}}, \bibinfo {author} {\bibfnamefont
  {L.-M.}\ \bibnamefont {Chen}}, \bibinfo {author} {\bibfnamefont
  {I.}~\bibnamefont {Daito}}, \bibinfo {author} {\bibfnamefont
  {K.}~\bibnamefont {Ogura}}, \bibinfo {author} {\bibfnamefont
  {T.}~\bibnamefont {Homma}}, \bibinfo {author} {\bibfnamefont
  {Y.}~\bibnamefont {Hayashi}}, \bibinfo {author} {\bibnamefont {H.Kotaki}},
  \bibinfo {author} {\bibfnamefont {A.}~\bibnamefont {Sagisaka}}, \bibinfo
  {author} {\bibfnamefont {M.}~\bibnamefont {Mori}}, \bibinfo {author}
  {\bibfnamefont {J.~K.}\ \bibnamefont {Koga}}, \bibinfo {author}
  {\bibfnamefont {T.}~\bibnamefont {Kawachi}}, \bibinfo {author} {\bibfnamefont
  {H.}~\bibnamefont {Daido}}, \bibinfo {author} {\bibfnamefont {S.~V.}\
  \bibnamefont {Bulanov}}, \bibinfo {author} {\bibfnamefont {T.}~\bibnamefont
  {Kimura}}, \bibinfo {author} {\bibfnamefont {Y.}~\bibnamefont {Kato}},\ and\
  \bibinfo {author} {\bibfnamefont {T.}~\bibnamefont {Tajima}},\ }\href@noop {}
  {\bibfield  {journal} {\bibinfo  {journal} {Phys. Plasma}\ }\textbf {\bibinfo
  {volume} {14}},\ \bibinfo {pages} {123106} (\bibinfo {year}
  {2007})}\BibitemShut {NoStop}%
\bibitem [{\citenamefont {Bulanov}\ and\ \citenamefont
  {Sakharov}(1991)}]{Sakharov1}%
  \BibitemOpen
  \bibfield  {author} {\bibinfo {author} {\bibfnamefont {S.~V.}\ \bibnamefont
  {Bulanov}}\ and\ \bibinfo {author} {\bibfnamefont {A.~S.}\ \bibnamefont
  {Sakharov}},\ }\href@noop {} {\bibfield  {journal} {\bibinfo  {journal} {JETP
  Lett.}\ }\textbf {\bibinfo {volume} {54}},\ \bibinfo {pages} {208} (\bibinfo
  {year} {1991})}\BibitemShut {NoStop}%
\bibitem [{\citenamefont {Matlis}\ \emph {et~al.}(2006)\citenamefont {Matlis},
  \citenamefont {Reed}, \citenamefont {Bulanov}, \citenamefont {Chvykov},
  \citenamefont {Kalintchenko}, \citenamefont {Matsuoka}, \citenamefont
  {Rousseau}, \citenamefont {Yanovsky}, \citenamefont {Maksimchuk},
  \citenamefont {Kalmykov}, \citenamefont {Shvets},\ and\ \citenamefont
  {Downer}}]{Matlis1}%
  \BibitemOpen
  \bibfield  {author} {\bibinfo {author} {\bibfnamefont {N.~H.}\ \bibnamefont
  {Matlis}}, \bibinfo {author} {\bibfnamefont {S.}~\bibnamefont {Reed}},
  \bibinfo {author} {\bibfnamefont {S.~S.}\ \bibnamefont {Bulanov}}, \bibinfo
  {author} {\bibfnamefont {V.}~\bibnamefont {Chvykov}}, \bibinfo {author}
  {\bibfnamefont {G.}~\bibnamefont {Kalintchenko}}, \bibinfo {author}
  {\bibfnamefont {T.}~\bibnamefont {Matsuoka}}, \bibinfo {author}
  {\bibfnamefont {P.}~\bibnamefont {Rousseau}}, \bibinfo {author}
  {\bibfnamefont {V.}~\bibnamefont {Yanovsky}}, \bibinfo {author}
  {\bibfnamefont {A.}~\bibnamefont {Maksimchuk}}, \bibinfo {author}
  {\bibfnamefont {S.}~\bibnamefont {Kalmykov}}, \bibinfo {author}
  {\bibfnamefont {G.}~\bibnamefont {Shvets}},\ and\ \bibinfo {author}
  {\bibfnamefont {M.~C.}\ \bibnamefont {Downer}},\ }\href@noop {} {\bibfield
  {journal} {\bibinfo  {journal} {Nat. Phys.}\ }\textbf {\bibinfo {volume}
  {2}},\ \bibinfo {pages} {749} (\bibinfo {year} {2006})}\BibitemShut {NoStop}%
\bibitem [{\citenamefont {Gradshteyn}\ and\ \citenamefont
  {Ryzhik}(2007)}]{Gradshteyn1}%
  \BibitemOpen
  \bibfield  {author} {\bibinfo {author} {\bibfnamefont {I.}~\bibnamefont
  {Gradshteyn}}\ and\ \bibinfo {author} {\bibfnamefont {I.}~\bibnamefont
  {Ryzhik}},\ }\href@noop {} {\emph {\bibinfo {title} {Table of Integrals,
  Series, and Products}}}\ (\bibinfo  {publisher} {Elsevier},\ \bibinfo {year}
  {2007})\BibitemShut {NoStop}%
\bibitem [{\citenamefont {Valenta}\ \emph {et~al.}(2020)\citenamefont
  {Valenta}, \citenamefont {Esirkepov}, \citenamefont {Koga}, \citenamefont
  {Pirozhkov}, \citenamefont {Kando}, \citenamefont {Kawachi}, \citenamefont
  {Liu}, \citenamefont {Fang}, \citenamefont {Chen}, \citenamefont {Mu},
  \citenamefont {Korn}, \citenamefont {Klimo},\ and\ \citenamefont
  {Bulanov}}]{Valenta1}%
  \BibitemOpen
  \bibfield  {author} {\bibinfo {author} {\bibfnamefont {P.}~\bibnamefont
  {Valenta}}, \bibinfo {author} {\bibfnamefont {T.~Zh.}\ \bibnamefont
  {Esirkepov}}, \bibinfo {author} {\bibfnamefont {J.~K.}\ \bibnamefont {Koga}},
  \bibinfo {author} {\bibfnamefont {A.~S.}\ \bibnamefont {Pirozhkov}}, \bibinfo
  {author} {\bibfnamefont {M.}~\bibnamefont {Kando}}, \bibinfo {author}
  {\bibfnamefont {T.}~\bibnamefont {Kawachi}}, \bibinfo {author} {\bibfnamefont
  {Y.-K.}\ \bibnamefont {Liu}}, \bibinfo {author} {\bibfnamefont
  {P.}~\bibnamefont {Fang}}, \bibinfo {author} {\bibfnamefont {P.}~\bibnamefont
  {Chen}}, \bibinfo {author} {\bibfnamefont {J.}~\bibnamefont {Mu}}, \bibinfo
  {author} {\bibfnamefont {G.}~\bibnamefont {Korn}}, \bibinfo {author}
  {\bibfnamefont {O.}~\bibnamefont {Klimo}},\ and\ \bibinfo {author}
  {\bibfnamefont {S.~V.}\ \bibnamefont {Bulanov}},\ }\href@noop {} {\bibfield
  {journal} {\bibinfo  {journal} {Phys. Plasma}\ }\textbf {\bibinfo {volume}
  {27}},\ \bibinfo {pages} {032109} (\bibinfo {year} {2020})}\BibitemShut
  {NoStop}%
\bibitem [{\citenamefont {Kulagin}\ \emph {et~al.}(2007)\citenamefont
  {Kulagin}, \citenamefont {Cherepenin}, \citenamefont {Hur},\ and\
  \citenamefont {Suk}}]{Kulagin1}%
  \BibitemOpen
  \bibfield  {author} {\bibinfo {author} {\bibfnamefont {V.~V.}\ \bibnamefont
  {Kulagin}}, \bibinfo {author} {\bibfnamefont {V.~A.}\ \bibnamefont
  {Cherepenin}}, \bibinfo {author} {\bibfnamefont {M.~S.}\ \bibnamefont
  {Hur}},\ and\ \bibinfo {author} {\bibfnamefont {H.}~\bibnamefont {Suk}},\
  }\href@noop {} {\bibfield  {journal} {\bibinfo  {journal} {Phys. Plasma}\
  }\textbf {\bibinfo {volume} {14}},\ \bibinfo {pages} {113101} (\bibinfo
  {year} {2007})}\BibitemShut {NoStop}%
\bibitem [{\citenamefont {Esirkepov}\ \emph {et~al.}(2009)\citenamefont
  {Esirkepov}, \citenamefont {Bulanov}, \citenamefont {Kando}, \citenamefont
  {Pirozhkov},\ and\ \citenamefont {Zhidkov}}]{Esirkepov2}%
  \BibitemOpen
  \bibfield  {author} {\bibinfo {author} {\bibfnamefont {T.~Zh.}\ \bibnamefont
  {Esirkepov}}, \bibinfo {author} {\bibfnamefont {S.~V.}~\bibnamefont {Bulanov}},
  \bibinfo {author} {\bibfnamefont {M.}~\bibnamefont {Kando}}, \bibinfo
  {author} {\bibfnamefont {A.~S.}\ \bibnamefont {Pirozhkov}},\ and\ \bibinfo
  {author} {\bibfnamefont {A.~G.}\ \bibnamefont {Zhidkov}},\ }\href@noop {}
  {\bibfield  {journal} {\bibinfo  {journal} {Phys. Rev. Lett}\ }\textbf
  {\bibinfo {volume} {103}},\ \bibinfo {pages} {025002} (\bibinfo {year}
  {2009})}\BibitemShut {NoStop}%
\bibitem [{\citenamefont {Berestetskii}\ \emph {et~al.}(1982)\citenamefont
  {Berestetskii}, \citenamefont {Lifshitz},\ and\ \citenamefont
  {Pitaevskii}}]{Lifshitz1}%
  \BibitemOpen
  \bibfield  {author} {\bibinfo {author} {\bibfnamefont {V.~B.}\ \bibnamefont
  {Berestetskii}}, \bibinfo {author} {\bibfnamefont {E.~M.}\ \bibnamefont
  {Lifshitz}},\ and\ \bibinfo {author} {\bibfnamefont {L.~P.}\ \bibnamefont
  {Pitaevskii}},\ }\href@noop {} {\emph {\bibinfo {title} {Quantum
  Electrodynamics}}}\ (\bibinfo  {publisher} {Pergamon},\ \bibinfo {year}
  {1982})\BibitemShut {NoStop}%
\bibitem [{\citenamefont {Narozhny}\ \emph {et~al.}(2004)\citenamefont
  {Narozhny}, \citenamefont {Bulanov}, \citenamefont {Mur},\ and\ \citenamefont
  {Popov}}]{Narozhny1}%
  \BibitemOpen
  \bibfield  {author} {\bibinfo {author} {\bibfnamefont {N.~B.}\ \bibnamefont
  {Narozhny}}, \bibinfo {author} {\bibfnamefont {S.~S.}\ \bibnamefont
  {Bulanov}}, \bibinfo {author} {\bibfnamefont {V.~D.}\ \bibnamefont {Mur}},\
  and\ \bibinfo {author} {\bibfnamefont {V.~S.}\ \bibnamefont {Popov}},\
  }\href@noop {} {\bibfield  {journal} {\bibinfo  {journal} {Phys. Lett. A}\
  }\textbf {\bibinfo {volume} {330}},\ \bibinfo {pages} {1} (\bibinfo {year}
  {2004})}\BibitemShut {NoStop}%
\bibitem [{\citenamefont {Aleksandrov}\ \emph {et~al.}(2019)\citenamefont
  {Aleksandrov}, \citenamefont {Plunien},\ and\ \citenamefont
  {Shabaev}}]{Aleksandrov1}%
  \BibitemOpen
  \bibfield  {author} {\bibinfo {author} {\bibfnamefont {I.~A.}\ \bibnamefont
  {Aleksandrov}}, \bibinfo {author} {\bibfnamefont {G.}~\bibnamefont
  {Plunien}},\ and\ \bibinfo {author} {\bibfnamefont {V.~M.}\ \bibnamefont
  {Shabaev}},\ }\href@noop {} {\bibfield  {journal} {\bibinfo  {journal} {Phys.
  Rev. D}\ }\textbf {\bibinfo {volume} {99}},\ \bibinfo {pages} {016020}
  (\bibinfo {year} {2019})}\BibitemShut {NoStop}%
\end{thebibliography}
\end{document}